\documentclass[12pt]{article}
\usepackage{amsfonts,bbold}
\usepackage[mathscr]{eucal}

\begin{document}

\begin{flushright}
  \small{UdeA-PE-002\\[3pt]
  \bf{hep-th/9906028}}
\end{flushright}

\vspace{2cm}

\begin{center}
  \Large{Regularization of  Automorphic Functions of\\
    Manifolds with Special K\"ahler Geometry. }

\vspace{1.5cm}

\large{Nelson Vanegas-Arbel\'aez} \\[5pt] 
\small{Departmento de F\'{\i}sica\\
Universidad de Antioquia.\\
A.A. 1226, Medell\'{\i}n, Colombia.\\
\textit{nvanegas@axion.udea.edu.co}} 
\end{center}

\vspace{3cm}

\abstract{In this paper we find automorphic functions of coset manifolds
  with special K\"ahler geometry. We use $\zeta$-functions to regularize an
  infinite product over integers which belong to a duality-invariant
  lattice, this product is known to produce duality-invariant functions.
  In turn these functions correspond to Eisenstein series which can be
  understood as string theory amplitudes that receive contributions from BPS
  states. The Ansatz is constructed using the coset manifold $SU(1,n)\over
  SU(n) \times U(1)$ as an example but it can be generalized.  Automorphic
  functions play an important role in the calculation of threshold
  corrections to gauge coupling and other stringy phenomena. We also find
  some connections between the theory of Abelian varieties and moduli
  spaces of Calabi-Yau manifolds. \vspace{1cm}}

\pagebreak

\section{Introduction}

\markright{\thesection. \quad Introduction.}
\label{section:intro}

It is known that special K\"ahler manifolds \cite{stro90,wit95} play an
important role in the low-energy effective actions derived from
superstrings. For example, the moduli space of of superstring theories
compactified on Calabi-Yau manifolds have special K\"ahler geometry
\cite{cand91-2,cand91,cand90}; this is also the type of manifold
parametrized by the scalar components of vector multiplets in $N=2$
supergravity theories in four dimensions \cite{zumi79}. 

Supergravity models based on the special K\"ahler manifold $SU(1,n)\over
SU(n) \times U(1)$ are known as ``minimal models''.  This is because the
manifolds $SU(1,n) \over SU(n) \times U(1)$ are the ones one gets
straightforwardly from superspace or superconformal tensor calculus
\cite{seib88}.  In this way this coset provides a somehow general setting
for supergravity models exhibiting special geometry. On the other hand,
this cosets are a non-compact analogous to $CP^n$ spaces with the avantage
of being the simplest family of Special K\"ahler manifolds \cite{ceco89}.

One expects the low-energy effective theories derived from superstrings to
be duality-invariant.  Any supergravity can be written in terms of only two
functions, namely $\mathcal{W}$ and $\mathcal{K}$, of the superfields.
They are known as the superpotential and the K\"ahler potential,
respectively (in turn this potential can be written as derivatives of
another function called the pre-potential).  These functions can actually
be combined in one expression,
\begin{equation}
  \label{eq:potentialG}
\mathcal{G} = \exp( \mathcal{K}) \, |\mathcal{W}|^2.  
\end{equation}
Duality transformations mix the moduli (the scalar field components of
$\mathcal{K}$ and $\mathcal{W}$) among themselves and leave the physics
unchanged. This means that K\"ahler potential transforms like
$$\mathcal{K} \rightarrow \mathcal{K} + \mathrm{holomorphic}
+\mathrm{anti-holomorphic}$$
under duality and therefore the superpotential
should transform like $\mathcal{W} \rightarrow e^{\lambda}\, \mathcal{W}$,
where $\lambda$ is some factor of automorphy, in order to keep
$\mathcal{G}$ invariant.



Duality transformation act on the K\"ahler potential. In general
$\mathcal{K}$ can be written as the symplectic product of two vectors (for
the class of manifolds under consideration); therefore, it is invariant
under symplectic transformations --up to a K\"ahler factor which does not
affect the K\"ahler metric. The duality group can be embedded into the
symplectic group, $\mathcal{D} \subset Sp(2n, \mathbb{Z})$ and, based on
this fact, one can construct an Ansatz to obtain duality-invariant
functions (in the case in which the moduli space has special K\"ahler
geometry). In fact one can propose
\begin{equation}
  \mathscr{G} = {\sum_{p_i,q_i}}' \, \log \, i \, \frac{| p_i t^i + q^i
    \mathcal{F}_i |^2}{t^i \bar{\mathcal{F}}_i -  \bar{t}^i \mathcal{F}_i}; 
\label{eq:master}
\end{equation}
with $\vec{p}= \{p_i\}, \, \vec{q}= \{q_i\} \in \mathbb{Z}^n, \quad \vec{p}
\neq \vec{0} \neq \vec{0}$, and the prime over sum indicating that it is to
be carried out for ``orbits'' of some duality-invariant lattice $\Gamma$
(we will see this later). The $t^i$ are the moduli and, as we will see
later in section \ref{section:su1}, $\mathcal{F}_i$ are derivatives of the
prepotential function $\mathcal{F}$, $\,\mathcal{F}_i = \partial_i
\mathcal{F}$. By construction this quantity has the required invariance.

Let us re-write $\mathscr{G}$ of eq.(\ref{eq:potentialG}) as
\begin{equation}
  e^{\mathscr{G}} = e^{K} \, |\Delta(t^i)|^2,
\end{equation}
formally one has that $\Delta$ is infinite and therefore needs to be
regularized (the factor of $1/2$ in front of $K$ is irrelevant given its
logarithmic form). From this expresion one obtains
\begin{equation}
  \Delta(t^i) = \prod_{\vec{p}, \vec{q} \in \Gamma} \, (p_i t^i + q^i
  \mathcal{F}_i). \label{eq:product}
\end{equation}
The suggested way for the regularization of this quantity is the so called
zeta-function regularization \cite{ferr91,ferr92}; this will be the object
of study in this paper.

We end this section by briefly discussing the threshold corrections to
gauge couplings in low-energy effective field theories derived from
superstrings \cite{give94}.  First one should notice that the formula
(\ref{eq:product}) has the same form of the threshold effect function
\cite{dixo90,dixo91,anto93}, which describes the one-loop string
corrections to the gauge coupling constant $g_a$ of the low energy
effective actions.  Indeed, this function has the form
\begin{equation}
 \label{eq:thres-corr}
  \mathscr{T}(\tau, \bar{\tau}) = \frac{1}{16 \pi^2}\, {C}_a \log \Big( 2
  \,\,   \mbox{Im}(\tau) \,\, |\eta(\tau)|^4 \Big)
\end{equation}
where $C_a$ are constants, that can be computed from the massless spectrum,
and $\eta(\tau)$ is the Dedekind $\eta$-function. Here, $|\Delta|^2$ would
be the analog of that function.  Threshold corrections for orbifold cases
are considered in reference \cite{mayr93}.

It is understood that the one-loop correction to the gauge coupling
constants have non-holomorphic parts (for a review on the subject, see
references \cite{kap94,kap94-2,der92} and references therein).  In the
effective field theory Lagrangian of an $N=1$ supersymmetric theory, the
fermionic terms are determined by the bosonic part \cite{dixo91}. Apart
from the condition that the bosonic fields span a K\"ahler manifold, the
functions 
\begin{equation}
  f_{ab} (\phi) = \Big( \frac{1}{g^2(\phi)} \Big)_{ab} - \frac{i
           \mathrm{\Theta}_{ab}(\phi)}{8 \pi^2} 
\end{equation}
must be holomorphic functions of the moduli $\phi$. $g$ and
$\mathrm{\Theta}$ are scalar field-dependent gauge coupling and vacuum
angle respectively.

At tree-level in string theory these functions can be written in terms of
the dilaton-axion field $S$:
\begin{equation}
   f_{ab} (\phi) = k_a \delta_{ab} \, S \, ,
\end{equation}
where $k$ is the level of the Kac-Moody algebra.  One-loop corrections to
this formula provide a useful mechanism for breaking supersymmetry.  At
one-loop the renormalized gauge couplings of the effective theory can be
expressed as
\begin{equation}
  g^{-2}_a (\mu) = k_a \, g^{-2}_{\mbox{\tiny string}} +\frac{1}{16
       \pi^2}\,b_a \log \, \frac{M^2_ {\mbox{\tiny string}}}{\mu^2}
       +  \mathscr{T}(\tau, \bar{\tau})
\end{equation}
where $\mu$ is the renormalization scale, $g^{-2}_{\mbox{\tiny string}}
\approx \mbox{Re} S$, $\,b_a$ are coefficients of the one-loop
$\beta$-function of the low-energy effective theory and $\mathscr{T}$ is
given in (\ref{eq:thres-corr}). 

The non-holomorphicity of the renormalized gauge coupling is however not 
necessarily a problem since it can interpreted as an effective coupling, a
purely low-energy effect \cite{kap94}. 

More recently a lot of work in the computation of the threshold corrections
in more realistic models has been done. In particular, there are some
computations of the effective gauge couplings in orbifolds with Wilson
lines done in reference \cite{mayr95,mayr94}. In these references the
threshold effect is given in terms of Siegel modular forms. 

Recent developments show the importance of automorphic functions in other
aspects of string theory \cite{ober99,kiri97,greg97,gutp97} as well. In
particular there has been a lot of progress in the search for functions
with the right invariances. However, most of these has been done for
toroidal compactifications of the string. In this paper we attempt to
calculate functions with the right automorphy properties for the case of
Calabi-Yau compactifications.


In section \ref{section-zeta} we present how the regularization of the
infinite product of eq.(\ref{eq:product}) can be done in terms of a
$\zeta$-function and we also construct such function in section
\ref{eq:myzeta}.  After this we proceed to obtain, in sections
\ref{section:su1} and \ref{section:known}, well known results of Dixon and
Kaplunovski \cite{dixo90,dixo91} for the $SU(1,1)/U(1)$ manifold, by doing
this we present our Ansatz for other (more complicated) cases. Section
\ref{section:su2} contains the first of such cases, namely the $SU(1,2)/
U(1) \times SU(2)$ manifold. We find that the automorphic function found
has two different contributions, depending on the solutions to a
Diophantine equation (\ref{eq:new-const}), we analyze them separately in
the next subsections and provide the $\zeta$-regularized result in
eq.(\ref{eq:finalmente}), this has the asymptotic expansion
(\ref{eq:asympt}).  In section \ref{section:symplectic} we show the action
of the symplectic group on the Riemann Theta functions found in previous
sections, this provides the mechanism by which duality acts on the
functions (in a way in which regularization should not be broken); in turn,
this gives us the interpretation of (\ref{eq:omega}) as the period matrix
of some super-torus, with a map between its moduli space (an Abelian Variety)
and the special K\"ahler manifold.

\section{The Mass Formula.}
\markright{\thesection. \quad The Mass Formula}
\label{section:mass}

We can explain equation (\ref{eq:master}) in a more consistent way; not just
by constructing it by hand but by looking at the mass formula and the free
energy of the particular low energy theory under consideration
\cite{ferr91,sabr96-2}. In field theory one can obtain the tree-level
free-energy for bosonic field using the path integral
\begin{equation}
  Z \equiv e^{F_B} = \int [\mathcal{D} \phi] \, \exp ( -\phi^{\dag}
  M^2_{\phi} \phi ) + \cdots \label{eq:free-nergy}
\end{equation}
where the integral should be carried our over the massive states of the
theory. Neglecting derivative terms one can use this formula as a
topological free energy of the bosonic compactified string. In the bosonic
case it will be interested in the contribution to the integral by those
states in the spectrum which are due to string compactification (that carry
momentum and winding number). The path integral (\ref{eq:free-nergy}) then
gives in general
\begin{equation}
  F = \log (\det M^{\dag} M).
\end{equation}

For $N=1$ supergravity for instance \cite{crem83,ferr91,sugrabook}, the
mass matrix has following form:

\begin{equation}
  M_{ij} = e^{\mathscr{G}/2}\, \mathscr{G}_{i\bar{m}}^{-1/2}\, [
  \mathscr{G}_{mn} + \mathscr{G}_m \mathscr{G}_n - \mathscr{G}_{mn\bar{k}} 
  \mathscr{G}^{-1}_{\bar{k}l} \mathscr{G}^l ]\,
  \mathscr{G}_{\bar{n}j}^{-1/2}; \label{eq:mass-matrix}
\end{equation}
with the subindex $i$ denoting $$\mathscr{G}_i \equiv \frac{\partial
  \mathscr{G}}{\partial \phi_i} \qquad \mathscr{G}^i \equiv \frac{\partial
  \mathscr{G}}{\partial \bar{\phi}_i} \qquad \mathscr{G}_i^j \equiv
\frac{\partial \mathscr{G}}{\partial \phi_j \partial \bar{\phi}_i} $$ 
and we lower and raise index with
$$(\mathscr{G}^{-1})^i_j \mathscr{G}^j_k = \delta^i_k.$$  

Supersymmetry is preserved when all the fields have vacuum expectation
values that are invariant under supersymmetry transformations, in
particular this means that we should have $$
e^{\mathscr{G}/2} (\mathscr{G}^{-1})^i_j)\mathscr{G}_i = 0$$ for $N=1$,
since the expression  in r.h.s is proportional to $\delta \Psi$, the
variation under supersymmetry of the fermionic fields. 

Then we can write (\ref{eq:mass-matrix}) as
\begin{equation}
  (M^{\dag} M)_{\bar{i}j} = e^{K}\,\, \mathscr{G}^{-1/2}_{\bar{i}k} \,\,
  \bar{\mathscr{W}}_{\bar{k}\bar{l}} \, \,(\mathscr{G}^{-1})^{\bar{l}m}
  \,\, \mathscr{W}_{\bar{m}n} \,\,  \mathscr{G}^{-1/2}_{\bar{n}j}.    
\end{equation} 
Using the form $K = -\log \, Y$, it one finds that \cite{ferr91}  
\begin{equation}
  \det (M^{\dag} M) = \det \, \frac{|\mathscr{W}|^2}{Y}.
\end{equation}
In the Ansatz provided above it is therefore possible to read the
superpotential from the mass formula.

Indeed, for the string theory case, the corresponding mass matrix has been
found for many cases, if the moduli space \cite{ferr91} is
\begin{equation}
  \mathcal{M} = \frac{SO(2,2)}{SO(2) \times  SO(2)}
\end{equation}
corresponding to some orbifold compactifications, the lattice momenta
$\Gamma (2,2)$ is spanned by 
vectors $p^{i_L},p^{i_R}$ which satisfy
\begin{equation}
  (p^{i_L})^2 - (p^{i_R})^2 = 2 \mathbb{Z} \label{eq:levelmatch}
\end{equation}
which is the conjugate to the condition satisfied by the moduli fields.
The free energy is then given by
\begin{equation}
  F = {\sum_{n,n',m,m'}}' \, \log \, \frac{|m+ n\, TU +i (m' U +
    n'T)|^2}{(T + \bar{T})(U + \bar{U})} \label{eq:sum}
\end{equation}
where the integers $(0,0,0,0) \neq (n,n',m,m') \in \mathbb{Z}^4$ do not run
over oscillator excitations and satisfy a stronger condition than that of
equation (\ref{eq:levelmatch}), namely the belong to the orbit described by
\begin{equation}
  (p^{i_L})^2 - (p^{i_R})^2 = m n + m' n' = 0. \label{eq:orbit1}
\end{equation}
With this condition one has that (\ref{eq:sum}) becomes the level-matching
condition.

The infinite sum (\ref{eq:sum}), subject to the condition
(\ref{eq:orbit1}), can be found using zeta-functions following the
procedure of references \cite{ferr91,oogu91}. Ferrara \textit{et. al.}
found that
\begin{equation}
  F|_{reg} = \log \, \Big( |\eta(T)|^4 |\eta(T)|^4 (T + \bar{T})(U +
  \bar{U}) \Big) \label{eq:automor}
\end{equation}
where $\eta(T)$ is the Dedekind function, defined as
\begin{equation}
  \eta(T) \equiv e^{\frac{1}{12} \pi T} \, \prod_{n>0} [1-e^{2 \pi n T}]. 
\end{equation}
However, the form of the actual calculation, is difficult to generalize to
other cases. The condition (\ref{eq:orbit1}) is resolved in two different
orbits: $n=0, \,\, n'=0$ and $n=0, \,\, m'=0$. This decouples
(\ref{eq:sum}) into two separate sums ($T^1 = T, \,\, T^2=U$) ,
\begin{equation}
  F(T_i) = \sum_{i=1}^{2} \sum_{(n,m) \neq (0,0)} \log \,\, \frac{m_i + i
    n_i T^i}{(T^i + \bar{T^i})}
\end{equation}
whose solution is (\ref{eq:automor}). We will do this calculation later in
this paper using a  different approach to that of reference
\cite{ferr91}.

The mass formula for other moduli spaces is also known to be of the form
(\ref{eq:master}).  For example, in references \cite{lope95,sabr96-2} it is
shown that, for a moduli space isomorphic to $\frac{SU(1,2)}{SU(2) \times
  U(1)}$, the chiral mass formula is of the form
\begin{equation}
  M^2 =\,\, 2 \,\,\frac{|m \mathcal{A} - n \tau + p|^2}{1- \tau \bar{\tau}
             - \mathcal{A}\bar{\mathcal{A}}} \label{eq:mass-su1n} \\[5pt]
\end{equation}
where $\tau, \mathcal{A}$ are the two complex moduli and the Gaussian
integers $m,n,p$ (integers of the form $a + ib$ with $a,b \in \mathbb{Z}$)
satisfy the level matching condition
\begin{equation}
  |m|^2 - |n|^2 - |p|^2 = 0.  \label{eq:orbit2} \\[5pt]
\end{equation}

This constraint is not easy to solve and therefore makes it difficult to
find a regularization for (\ref{eq:mass-su1n}). In \cite{lope95} a solution
is proposed for the automorphic function of $\frac{SU(1,n)}{SU(n) \times
  U(1)}$ by truncating a $\frac{S0(2,4)}{SO(2) \times SO(4)}$ coset. The
sums are carried out in orbits of the sublattice $\Gamma_p$ of the $E_8
\otimes E_8$ lattice $\Gamma_{16}$. The orbits considered not only obey the
level matching condition but also severely restrict the numbers over which
the sum takes place.  

\section{The Zeta Function.}
\markright{\thesection. \quad Zeta Function Techniques}
\label{section-zeta}

We briefly introduce the powerful mechanism of zeta-function regularization
\cite{quin93}. First consider a simple example
\cite{ramond,number} in field theory in which $\zeta$-functions are use to
find determinants.

Let us consider a quantum operator $\Lambda$ with eigenvalues $\lambda_n$,
it obeys the eigenvalue equation $ \Lambda f_n (x) = \lambda_n f_n (x)$. We
define the $\zeta$-function associated with $ \Lambda $ as
\begin{equation}
  \zeta_{\Lambda} (s)\, =\, \sum_{n=1}^{\infty}
         \,\,\Big(\frac{1}{\lambda_n} \Big)^s 
\end{equation}
if $\Lambda = \mathbb{1}$, this function would coincide with the Riemann
function. From this we can see that 
\begin{equation}
  \det \, \Lambda \equiv \prod_{n} \, \lambda_n = e^{- \zeta'(0)}.
\end{equation}

To find $\zeta_{\Lambda} (s)$, we try to solve the heat equation
\begin{equation}
  \Lambda_x \, G (x,y,t) = - \frac{\partial\phantom{t}}{\partial t} \,  
                           G(x,y,t), \label{eq:heateq}
\end{equation}
with the boundary condition $G(x,y,0) = \delta (x-y)$.  We construct the
heat function $G$ given as
\begin{equation}
  G (x,y,t) \equiv \sum_n \, e^{-\lambda_n t} \, f_n(x) f^*_n(y).
\end{equation}
It can easily be verified that $G$ solves (\ref{eq:heateq}).  The
associated  $\zeta$-function  can expressed as 
\begin{equation}
  \zeta_{\Lambda} (s)\, =\, \frac{1}{\Gamma(s)} \int_0^{\infty} \, dt
                         \, t^{s-1} \int d^p x \,  G (x,x,t).
\end{equation}
This analytic form of the  $\zeta$-function can then be used to compute a
regularized determinant of the operator $\Lambda$.

\subsection{Regularization of the Automorphic Function.}
\markright{\thesection. \quad Regularization of the Automorphic...}

With all this in mind, we can propose as an Ansatz for the regularization of
the infinite product (\ref{eq:product}), via the limit formula
\begin{eqnarray}
  \mathscr{G} & =  & \log \,\,(\,\, |\Delta|^2 \, e^{K} \,\,)\, 
     = \,\, - {\sum_{(p_i,q_i)\in L_{\Gamma}}} \, 
     \log \, i \, \frac{| p_i t^i + q^i  \mathcal{F}_i |^2}{(t^i
     \bar{\mathcal{F}}_i -  \bar{t}^i \mathcal{F}_i)}  \nonumber \\[6pt]
     & = & -\,\, \Big[\, {\sum_{\,(p_i,q_i)\in L_{\Gamma}}} \! [\, \log \,\,
     (p_i t^i + q^i \mathcal{F}_i) + \mbox{c.c.}\,] \,\,\, -
     \;{\sum_{(p_i,q_i)\in L_{\Gamma}}} \!\! \log \,\, (t^i
     \bar{\mathcal{F}}_i -  \bar{t}^i \mathcal{F}_i) \,\, \Big]
     \nonumber  \\[6pt] 
     & = & \, - \, \lim_{s \rightarrow 0}\,\, \frac{d\phantom{s}}{ds}
     \,\,\zeta(s)  . \label{eq:limit}
\end{eqnarray}
With the associated  $\zeta$-function  defined as
\begin{eqnarray}
  \zeta(s) & = &  \frac{1}{\Gamma(s)}\,\, \int_0^{\infty} \, d
               \tau \, 
              \tau^{s-1} \sum_{(p,q) \in L_{\Gamma}}\Big[ \, \exp \,
              \Big(- i \tau \,\pi ( p_i t^i + q^i  \mathcal{F}_i ) \Big)
              \, +  \nonumber  \\[6pt]
           & & \qquad \qquad \exp \,
              \Big(- i \tau \,\pi ( \bar{p}_i \bar{t}^i + \bar{q}^i
              \bar{\mathcal{F}}_i ) \Big) - \frac{1}{2}
              \exp \,\Big( - i \tau\, \pi (t^i \bar{\mathcal{F}}_i -
              \bar{t}^i \mathcal{F}_i) \Big) \Big] \,; \label{eq:myzeta}
\end{eqnarray}

\noindent the sums are taken over the orbits $L_{\Gamma}$ in the sublattice
spanned by the momentum and winding numbers, not considering oscillator
excitations. 

The integers $(q,p)$ obey constraints similar to those of equations
(\ref{eq:orbit1},\ref{eq:orbit2}), specialized for the particular example
of the special manifold  under consideration.  They transform as a
vector under the symplectic group of which the duality group is a subgroup.
This is easily seen since we can write the numerator of $\mathscr{G}$ as
\vspace{0mm}
\begin{equation} 
  \left(  
    \begin{array}{c}
      \vec{\bar{q}} \\ 
    i \vec{\bar{p}}
    \end{array}
    \right)^{\dag} 
    \left( 
\begin{array}{cc} 0 & \mathbb{1}  \\
           -\mathbb{1} & 0 
\end{array} \right) 
\left( \begin{array}{c} \vec{t} \\ 
                     i\vec{\mathcal{F}} 
       \end{array} \right); \\[7pt]
\end{equation}

\noindent since $(t, \mathcal{F})$ is a vector under $Sp(2n, \mathbb{R})$
we must have that, for $\mathscr{G}$ to be invariant under duality
transformations, $(\bar{p},\bar{q})$ must transform as a symplectic vector
as well. Moreover, if $\mathcal{G}(\mathbb{R})$ is the duality group then,
given the embedding of the duality group into the symplectic group,
$(\bar{p},\bar{q})$ should transform as a vector of
$\mathcal{G}(\mathbb{Z})$.

\section{The $SU(1,1)/ U(1)$ Case.}
\markright{\thesection. The $SU(1,1)/ U(1)$ Case.}
\label{section:su1}

We should now try to solve the first and most simple case of special
K\"ahler manifold. As mentioned earlier, this example is solved in
reference \cite{ferr91}. However, we are going to consider a different
approach here which will have the advantage of being more general and
therefore possible to apply to other cases.

The particular form of $\mathscr{G}$ for this special manifold depends on
the coset representatives chosen. Indeed, if one chooses an unbounded
realization of the coset, the representatives  obey the constraint
\begin{equation}
x_{0}^{\dag} x_{0} - X^{\dag} X = 1 \label{eq:constrt}.
\end{equation}
in homogeneous coordinates. The prepotential $\mathcal{F}$ is given by
\begin{equation}
F \,(X)= \frac{i}{2} \, \eta^{IJ} \, {X}_I {X}_J, \label{eq:the-prep}
\end{equation}
we have
\begin{eqnarray}
  (p_i t^i + q^i \mathcal{F}_i)& =& (p_0 x_0 + p_1  x_1 + i q_0 x_0 + i q_1
  x_1) \nonumber \\ 
   & =& P_0 x_0 + P_1 x_1  \qquad \mbox{for} \qquad P_i = p_i + i q_i
\label{eq:explicit-su1}
\end{eqnarray}
just as we expected from the form of the chiral mass formula
(\ref{eq:mass-su1n}). The integers are subject to certain conditions. In
fact they satisfy the constraint $|P_0|^2 - |P_1|^2 =0$, in the coset
manifold we are considering,
\begin{equation}
  (p_0)^2 + (q_0)^2 - (p_1)^2 - (q_1)^2 =0; \label{eq:int-const}
\end{equation}
the sum (\ref{eq:myzeta}) over $(p,q)$ is carried out only over these
$SU(1,1) \equiv L_{\Gamma} $ orbits.

The constraint (\ref{eq:int-const}) can be generally solved. However, we
first should impose an extra condition to avoid over-counting. Indeed, when
one looks at the relation between the parameters of the coset $SU(1,n)/U(1)
\times SU(n)$, given by equation (\ref{eq:constrt}), it is obvious that the
parameter $x_0$ is real by definition.  Therefore, since the
$\zeta$-function defined so far includes $x_0$ as a complex variable and is
treated as any other parameter $x_i$, we need to impose some condition to
avoid counting $x_0$ twice. The obvious choice is to impose a similar
constraint over the Gaussian integer associated with this parameter, we
then set $q_0 =0$, to set $P_0 \in \mathbb{R}$. 

The solution of the resulting constraint is an old number theory problem,
which has been considered by many mathematicians \cite{dickson}. The
proposed solution is
\begin{eqnarray}
  q_1 & =& 2\, n\, m  \nonumber \\ 
  p_1 & =& n^2 - m^2    \nonumber \\ 
  p_0 & =& n^2 + m^2  \qquad \mbox{for} \qquad (n,m) \in \mathbb{Z}^2.
\end{eqnarray}
The first term of the $\zeta$-function (\ref{eq:myzeta}) is therefore given
by
\begin{equation}
  \zeta (s) = \frac{1}{\Gamma(s)}\,\, \int_0^{\infty} d \tau \, 
         \tau^{s-1} \sum_{(m,n) \in \mathbb{Z}^2}  e^{-\tau  \pi 
         (n^2  (1+t) + m^2 (1+t) + 2 i n m t )  } \\[6pt] \label{eq:suma}
\end{equation}

\noindent where we have used the parametrization of $SU(1,1)/U(1)$ given by
the inhomogeneous physical coordinates $t = x_1$ and $x_0 =1$, and with the
condition $(n,m) \neq (0,0)$.  After this, our $\zeta$-function can be
written as
\begin{equation}
    \zeta (s) = \frac{1}{\Gamma(s)}\,\, \int_0^{\infty} d \tau \, 
              \tau^{s-1} \sum_{(m,n) \in \mathbb{Z}^2}  \exp [ -\tau \pi
              \mathbf{Q}(n,m,t)]
\label{eq:theta}
\end{equation}
where
\begin{eqnarray}
  \mathbf{Q} (n,m,t) & = & \left(
       \begin{array}{c} n\\m \end{array} \right)^t \cdot              
  \left(
  \begin{array}{cc}
    1 + t &   it  \\ i t & 1 -  t 
  \end{array}  \right) \cdot
  \left(
       \begin{array}{c} n\\m \end{array} \right) \nonumber \\[8pt] 
  & \equiv & \vec{n}^t \,\, \Omega \,\, \vec{n}  \label{eq:cuad} .
\end{eqnarray}
\noindent The matrix $\Omega$ satisfies
\begin{eqnarray}
  \det \, \Omega \,& = & \, 1  \nonumber  \\[7pt] 
   \Omega^{-1} \, & = & \,
        \,\, \left(  
    \begin{array}{cc}
       1-t & -i t\\
       -i t & 1+t
    \end{array} \right). \label{eq:inv-omega}
\end{eqnarray}

The sum over integers in (\ref{eq:theta}) is an example of the Riemann
Theta function \cite{abelian}, which we will briefly review in the next
section. But finally we should note that we could have constructed a
solution to (\ref{eq:int-const}) without setting $q_0 =0$. However, it
turns out that the matrix $\Omega$ in such solution in not invertible
(always has an eigenvalue zero). It seems that this condition is justified
in every sense.

\subsection{Riemann Theta Functions.}
\markright{\thesection. Riemann Theta Functions}
\label{riemanns}

Riemann Theta function is defined in general as
\begin{equation}
 \theta (z, \mathbb{\Omega}) \equiv \sum_{n \in \mathbb{Z}^g} \exp 
               \left(
               \pi i n^t \cdot \mathbb{\Omega} \cdot n + 2 \pi i n^t 
               \cdot z \right) \label{eq:riem-theta},
\end{equation}
where $z \in \mathbb{C}^g$, $\mathbb{\Omega}$ is a $p \times p$ complex
matrix which 
satisfies
\begin{equation}
  \mathbb{\Omega}^t = \mathbb{\Omega} \qquad \mbox{and} \qquad \mbox{Im} \,
  \mathbb{\Omega} > 0. 
\end{equation}
These conditions define the Siegel upper half space $\mathfrak{H}_g$. It is
a generalization to the usual upper complex plane $\mathfrak{H}_1$ defined
by Im $\tau > 0$, these conditions guarantees the convergence of
(\ref{eq:riem-theta}).  These conditions are just a particular form of the
Riemann Bilinear Conditions. 

Riemann Theta functions satisfy beautiful periodicity conditions:
\begin{equation}
  \theta ( z+m, \mathbb{\Omega}) = \theta(z,\mathbb{\Omega})
  \label{eq:theta-inv} 
\end{equation} and
\begin{equation}
  \theta(z+\mathbb{\Omega} \cdot m, \mathbb{\Omega}) = e^{-(\pi i m^t \cdot
    \mathbb{\Omega} \cdot m 
    + 2 \pi i m^t \cdot z)} \theta(z,\mathbb{\Omega}), \label{eq:theta-inv1}
\end{equation}
for $z \in \mathbb{C}^g, \quad m \in \mathbb{Z}^g$.  A generalized version
of the Riemann theta functions is the theta function with characteristics
$\Theta$ [\raisebox{-.6ex}{$\stackrel{a}{\scriptstyle{b}}$}], defined as
\begin{eqnarray}
  \Theta [\raisebox{-.6ex}{$\stackrel{a}{\scriptstyle{b}}$}]
  (z,\mathbb{\Omega})       & = & \sum_{n
    \in \mathbb{Z}^p} \exp \, [\, \pi i \,(n+a)^t \cdot \mathbb{\Omega}
     \cdot (n+a) +   2 \pi i (n+a) \cdot (z+b)  ] \nonumber  \\[6pt] 
     & = &  \exp \, [\,\pi i a \cdot \mathbb{\Omega} \cdot a + 2 \pi i
     a^t \cdot 
     (z+b)] \,\, \Theta (z+ \mathbb{\Omega} \cdot a + b, \mathbb{\Omega});
     \label{eq:character} 
\end{eqnarray}
for $a,b \in \mathbb{Q}$. 

With this notation we now realize that (\ref{eq:theta}) can be written in
terms of the Riemann theta function $\theta(0,\Omega)$ as
\begin{equation}
  \zeta(s) = \frac{1}{\Gamma(s)}\,\, \int_0^{\infty} d \tau \, 
              \tau^{s-1} \Big(\theta_{\tau}(0,\Omega') -1 \Big) \\[6pt]
\label{eq:mellin}
\end{equation}
where $\theta_{\tau}(0,\Omega') \equiv \theta (0,i \tau \, \Omega)$. In
other words our $\zeta$-function is nothing but a Mellin transformation of
a Riemann theta function. The factor of $1$ correspond to the $0$-modes
that are not present in the original (\ref{eq:theta}). How
(\ref{eq:mellin}) transforms under duality will be discussed later.
 
The limit (\ref{eq:limit}) can be taken simply by considering the behavior
of $\Gamma (s)$ as $s$ goes to zero. If we denote (\ref{eq:mellin}) as
$\zeta (s) = (\Gamma (s))^{-1} \, I(s)$, we have 
\begin{equation}
  \lim_{s \rightarrow 0}\,\, \frac{d\phantom{s}}{ds}\,\,\zeta(s) =
       \lim_{s \rightarrow 0}\,\,\Big[\,I(s)
       \,\frac{d\phantom{s}}{ds}(\Gamma   (s))^{-1} 
        +  \frac{1}{\Gamma (s)} \, \frac{d\phantom{s}}{ds} I(s) \,
        \,\Big]. \\[6pt] \label{eq:deriv}
\end{equation}
The first term is easily solved, we make  use of \cite{ryder} 
\begin{eqnarray}
  \Gamma (\epsilon) & = & \frac{1}{\epsilon} - \gamma +
  \mathcal{O}(\epsilon^2) \nonumber \\
   \psi (\epsilon) & = & -\gamma -  \frac{1}{\epsilon} +
                         \mathcal{O}(\epsilon^2)   
         \quad  \mbox{with} \quad \psi (s)  \equiv   \frac{ d \, \ln
         \Gamma(s)}{ds}  = \frac{\Gamma'(s)}{\Gamma(s)}
\end{eqnarray}
where $\gamma =0.5772157... $ is the Euler-Mascheroni constant. Therefore
we have that 
$$\lim_{s \rightarrow 0}\,\, \frac{d}{ds}\, \frac{1}{\Gamma(s)} \approx
-1.$$ 

The second term in (\ref{eq:deriv}) is more involved.  We will prove that
$I(s)$ can be expanded in a Laurent series around $s \rightarrow 0$ with
simple pole at $s=0$. We will also calculate the residue and therefore we
should be able to write 
\begin{equation}
   I(s) \approx \frac{C_{-1}}{s} + C_0 +  \mathcal{O}(s) 
\label{eq:laurent}
\end{equation}
taking the derivative and then multiplying by $\frac{1}{\Gamma(s)} \approx
s$ for $s$ small, we can write 
\begin{eqnarray}
 \Psi \, (\,\Omega(t)) & \equiv &
   \lim_{s \rightarrow 0}\,\, \frac{d\phantom{s}}{ds}\,\,\zeta(s) \nonumber
   \\[6pt]  
     & = & - \lim_{s \rightarrow 0}\,\,\Big[ \int_0^{\infty} d \tau \,
     \tau^{s-1}  \Big(\theta_{\tau}(0,\Omega') -1 \Big) +
     \frac{C_{-1}}{s} \Big];
  \label{eq:kron}
\end{eqnarray}
hence the function $\Psi (\Omega)$ is well behaved and finite once the pole
is subtracted.

Invariances of (\ref{eq:kron}) are not difficult to establish. One
just has to look at the periodicity conditions (\ref{eq:theta-inv} and
\ref{eq:theta-inv1}) of the Riemann theta function for $z=0$. We will
discuss this bellow, first we need to establish some properties of
(\ref{eq:kron}). 

\subsection{Mellin Transformation.}
\markright{\thesection. Mellin Transformation}

\label{mellin}
The function $\Psi \,\, (\,\Omega(t), s) $ of equation (\ref{eq:kron}) is
known as a Mellin Transformation, we briefly review a few facts about this
transform \cite{number} since they are important in our study of the
automorphic functions.

We define the Mellin Transformation of a function $f(x)$ as
\begin{equation}
  \mathbf{M} \,[f(u)] \,\, \equiv \,\, \int_0^{\infty} d x\, x^{u-1} f(x);
\end{equation}
for instance the Gamma function, $\Gamma(s)$ is the Mellin Transformation
of the exponential function, $f(x) = e^{-x}$. The condition for $
\mathbf{M} \,[f(u)]$ to be properly defined is that, for $ x>0$, the
function $f(x)$ decays rapidly at infinity, and at the origin it diverges
at most as a polynomial $f(1/x) \sim \mathcal{O} (x^{-\mathrm{K}})$ as $x
\rightarrow 0$. With these conditions $ \mathbf{M}\,[f(u)]$ converges for
$u \in \mathbb{C}$ and $\mathrm{Re}\, s > \mathrm{K}$.

With the conditions described above we can study (\ref{eq:kron}). Lets
assume that a function $\phi(\tau)$, convergent as $\tau \rightarrow
\infty$, satisfies the functional equation
\begin{equation}
  \phi(\frac{1}{\tau}) = \sum_{k=1}^{\mathrm{K}} \, b_k \, \tau^{\lambda_k} +
                          \alpha \,\tau^h \, \phi(\tau)
                          \label{eq:functional} 
\end{equation}
for $\tau>0$ and $\alpha \neq 0,\,\, \lambda_k,\, h,\, b_k$ complex
numbers.  Then the Mellin transformation of $\phi(\tau)$ can be written as
a sum $\int_{0}^{\infty} = \int_0^1 + \int_1^{\infty}$, and we have
\begin{equation}
  \mathbf{M} \, [ \phi(\alpha, s)] = \int_1^{\infty} \Big[\sum_k b_k
                      \tau^{\lambda_k} 
                      + \alpha \,\tau^h \, \phi(\tau)\Big] \tau^{-s-1}
                      \,d\tau +  \int_1^{\infty} \phi(\tau)
                      \tau^{s-1} \, d \tau \\[6pt]
\end{equation}
where we have used the functional equation in the first term. 

After we perform the first of these integrals we are left with
\begin{equation}
  \mathbf{M} \, [ \phi(\alpha, s)]\, =\, \sum_k \frac{b_k}{s-\lambda_k}\, +
                \, \int_1^{\infty} \phi(\tau) \tau^{s-1} \, d\tau \, + \,
                \alpha \, \int_1^{\infty}  \phi(\tau) \tau^{h-s-1} \, d\tau
                . \\[7pt]
\end{equation}

The first term gives the poles of the Mellin transformation at $\lambda_k$
with residues $b_k$. The second term is convergent for all values of $s$.
If we now use the functional equation twice, we obtain a second version of
it
\begin{equation}
   \phi(\frac{1}{\tau}) =  \, \frac{1}{\alpha} \, \tau^h \,\phi(\tau) -
                         \frac{1}{\alpha} \sum_{k=1}^{\mathrm{K}} b_k
                         t^{h-\lambda_k}. 
\end{equation}

Integrating as before, we obtain 
\begin{equation}
  \mathbf{M} \, [\phi(\alpha, s)] \, =\, \sum_k
                \frac{-b_k/\alpha}{s- h + \lambda_k}\, + 
                \, \int_1^{\infty} \phi(\tau) \tau^{s-1} \, d\tau \, + \,
                \frac{1}{\alpha} \, \int_1^{\infty}  \phi(\tau)
                \tau^{h-s-1} \, d\tau .                
\end{equation}

Therefore we arrive at the conclusion that the Mellin transformation of
$\phi$ as defined, is invariant (up to a factor of $\alpha$) under $s
\rightarrow h-s, \,\, \alpha \rightarrow \frac{1}{\alpha}$,
\begin{equation}  
 \mathbf{M} \, [\,\phi(\alpha, s)\,] \, =\, \alpha\, \mathbf{M} \,
                                 [\,\phi(\frac{1}{\alpha}\, ,\, h- s)\,]; 
\end{equation}
with poles at $\lambda_k$ and residues  $b_k$.

In the Mellin transformation under consideration (\ref{eq:kron}), the
functional equation can be obtained using the Poisson's summation formula:
\begin{equation}
  \sum_{\vec{n} \in \mathbb{Z}^p} \, \exp [ -\pi \, \vec{n}^t \cdot A 
                     \cdot \vec{n}]\, =  \, (\det A)^{-\frac{1}{2}} 
                     \sum_{\vec{m} \in \mathbb{Z}^p} \, \exp [ -\pi \,
                     \vec{m}^t \cdot A^{-1} \cdot \vec{m}]. \\[6pt]
                     \label{eq:poisson} 
\end{equation}
\noindent This amounts to obtaining the transformations properties of
$\,\,\theta(0, \tau \Omega), \,\,$ defined by equation (\ref{eq:mellin}),
when $\tau \rightarrow \frac{1}{\tau}$.  

We define $\phi(\tau) \equiv \theta_\tau(0,\Omega) -1 \equiv \theta( \tau\,
\Omega) -1 $ and make use of (\ref{eq:poisson}) to get
\begin{eqnarray}
 \phi( \frac{1}{\tau}) &= &\theta(0,\frac{1}{\tau}\, \Omega)-1 
                           \nonumber \\[6pt]
           &= &  \frac{1}{(\det \tau^{-1}\, \Omega)^{\frac{1}{2}}} 
           \,\,\, \theta (0, \tau\, \Omega^{-1}) -1 \nonumber \\[7pt] 
           &= & \frac{\tau}{\alpha} \,\,\, \theta(\tau\, \Omega^{-1})-1 
\end{eqnarray}
where we have set $\,\alpha = (\det \,\Omega)^{1/2} \,$. However we
also have that
\begin{equation}
  \theta (\tau\, \Omega) = \theta (\tau \, \Omega^{-1}) \label{eq:thetainv}
\end{equation}
this is because the sum implied in the right hand side of the equation can
be obtained by re-arranging the sum in the left hand side. In fact, just by
looking at the form of $\Omega$ and $\Omega^{-1}$ in (\ref{eq:inv-omega})
and (\ref{eq:cuad}) one can see that, by redefining the integers in the sum
(\ref{eq:suma}) as
\begin{equation}
  n \rightarrow -m \qquad \mbox{and} \qquad m \rightarrow n,
\end{equation}
we go from the l.h.s. to the r.h.s. of (\ref{eq:thetainv}).

Therefore we have the following functional equation:
\begin{equation}
  \phi(\frac{1}{\tau})\, = \, \alpha \, \tau \,\phi(\tau)-1 + \alpha \,
  \tau ;
\end{equation}
where we just have to use what we have already learned about the Mellin
transformation to learn about $\Psi(\Omega)$ in (\ref{eq:kron}). We have,
in (\ref{eq:functional}),
\begin{eqnarray}
  h=1 \qquad  \mbox{and}& \qquad& \lambda_1 = 0 \qquad b_0 = -1   \nonumber 
 \\
  & \qquad & \lambda_2 = 1 \qquad b_2 = \alpha.
\end{eqnarray}

We conclude that $\Psi (s,\Omega)$ extends to a holomorphic function of $s$,
except for simple poles located at $s=0$ and $s=1$, with residues $-1$ and
$\,\alpha = (\det \,\Omega)^{1/2} \,$ respectively. Moreover, up to factors
of $\alpha$, the function $\Psi (s)$ is invariant under $s \rightarrow 1-s$.

Thus we have completed the expression (\ref{eq:kron}) which, after a mild
abuse of our notation, we write as
\begin{equation}
  \lim_{s \rightarrow 0} \,( \Psi (s, \Omega) + \frac{1}{s}) 
\end{equation}
the desired $\zeta$-function is finite and well defined. It is however
difficult to give an general solution to the Mellin transformation, we will
try to give an answer to that later.

\subsection{Putting It All Together.}
\markright{\thesection. Putting It All Together}

Having found the form of the first term in the expression (\ref{eq:myzeta})
it is now easy to compute the automorphic function of $\frac{SU(1,n)}{SU(n)
  \times U(1)}$. The second term is nothing but $\Psi (\Omega (\bar{t}))$,
therefore the first two terms pick the real part of this function.  The
third term can be computed using Riemann $\zeta$-functions as defined
before in section \ref{section-zeta}.

We recall here the $\zeta$-function related to the third term, it is given
by 
\begin{equation}
  \zeta_3(s) =  - \frac{1}{2} \,\, \frac{1}{\Gamma(s)}\,\, \int_0^{\infty}
              \, d \tau \,  
              \tau^{s-1} \,\, \sum_{0 \neq m,n \in \mathbb{Z}}
              \exp \, [ - \tau\, \pi (1- t\bar{t}) ].
\end{equation}

The contribution of this term to the function (\ref{eq:potentialG}) is
given by its derivative and in the limit $s \rightarrow 0$:
\begin{equation}
- \lim_{s \rightarrow 0} \,\, \frac{d\phantom{s}}{ds} \,\, \zeta_3(s) = 
   \frac{1}{2}
   \lim_{s \rightarrow 0} \,\, \frac{d\phantom{s}}{ds} \,\,\, 
   \Big[ \,\,  \frac{1}{\Gamma(s)}\,\,
   \int_0^{\infty} \, d \tau' \, \tau'^{s-1} e^{-\tau'} \,\,
   \sum_{0 \neq m,n \in \mathbb{Z}} \frac{1}{[ 1-t\bar{t} ]^s}
  \Big]
\end{equation}
the integral is the Gamma function, so we get
\begin{equation}
-  \lim_{s \rightarrow 0} \,\, \frac{d\phantom{s}}{ds} \,\, \zeta_3(s) = 
   \frac{1}{2}
   \lim_{s \rightarrow 0} \,\, \frac{d\phantom{s}}{ds} \,\,\, \Big[
   \, (1 - t \bar{t}) \,\, \Big]^{-s} \,\,\Big( 4 \, 
   \sum_{n>0 \,\,m>0} 1  \Big);
\end{equation}
here we use  $\zeta(0) = -1/2$, to finally get
\begin{equation}
  -  \lim_{s \rightarrow 0}\,\, \frac{d\phantom{s}}{ds} \,\, \zeta_3(s) 
     \, = \,\, \ln \,\, |1 - t\bar{t}|
\end{equation}

Collecting all the results of this section we can write 
\begin{equation}
  \label{eq:automorphic}
    e^{\mathscr{G}}  =   |\Delta|^2 \, e^{K} = \,\, 
    e^{ 4  \,\mathrm{Re}\, \Psi \,(\,\Omega(t)\,) }
       \,\, |1-t\bar{t}|
\end{equation}

Finding a form for this expression in terms of known functions of
mathematical physics is rather complicated.  We will make some attempts in
that direction later. It is also important to verify that the expression
found is invariant under duality transformation. The original expression
(\ref{eq:limit}) obviously is, but we should discuss this later with more
detail.

\section{Known Results.}
\markright{\thesection. Known Results}
\label{section:known}

We should now focus in the problem of finding an analytic form of the
function we have defined.  Fortunately this is possible for the case of the
coset we have just seen in the previous section and we should show that it
is possible in general.  In particular, we shall find the results of
Ferrara \textit{et. al.} in \cite{ferr91}.

To proceed we need to make a direct integration of the $\zeta$-function, as
defined in (\ref{eq:myzeta}). The first term renders the Eisenstein series
\begin{equation}
 \frac{1}{\Gamma(s)}  \int_0^{\infty} d \tau \,\tau^{s-1}
     \Big(\theta_{\tau}(0,\Omega') -1   \Big) = \sum_{m,n \neq 0} (\pi
     \mathbf{Q}(m,n))^{-s} \label{eq:eisenstein}
\end{equation}
where $\mathbf{Q}$ is the quadratic form defined (\ref{eq:cuad}) and where
we have also identified it with the Mellin transformation studied in the
preceding section.  This Eisenstein series is convergent for Re$s > 1$ but
can be analytically continued to the whole complex plane. The derivative of
the resulting function --just as we saw -- has poles at $s=0,1$. Taking
into account the three terms of the $\zeta$-function (\ref{eq:myzeta}) and
once the derivative against $s$ is taken, one can easily see that final
Eisenstein series to calculate becomes
\begin{equation}
  \mathbf{G} \, (t,s) = \sum_{m,n \neq (0,0)} \frac{( 1 - t 
    \bar{t})^{s} }{|\,\pi \, \mathbf{Q} \, (t,m,n)
    \,|^{2s}}. \label{eq:dual-form} 
\end{equation}

This form is important because it allow us to study duality transformations
on the the automorphic function in a convenient way. But, for the effects of
the calculation, we will work with the form broken up in three pieces and
will concentrate our attention in (\ref{eq:eisenstein}).

First we will change our variables as follows:
\begin{equation}
  t \rightarrow T \,\, = \,\, \frac{1-t}{t+1} \quad \mbox{and} \quad t=
  \frac{1-T}{T+1} 
\end{equation}
which means, for the automorphic function,
\begin{eqnarray}
  \Omega  & \rightarrow &  \mathbf{\Omega} \,\, = \,\, \left( 
    \begin{array}{cc}
      1 & \frac{i}{2} \,\, (1-T)\\
      \frac{i}{2} \, (1-T) & T 
    \end{array} \right) \nonumber \\[6pt]
  (1- t \bar{t}) & \rightarrow & \frac{1}{2} \,\,( T +
  \bar{T}). \label{eq:newq} 
\end{eqnarray}

Using the so-called Chowla-Selberg formula \cite{chow49}, we can find an
analytic form of the automorphic function, in terms of known functions of
mathematical physics. This process is similar to the demonstration of
Kronecker's First Limit Formula.  To do this we need to use the general form
of the Poisson summation formula, namely,
\begin{equation}
  \sum_{n \in \mathbb{Z}} \,\, g (x +n) = \,\, \sum_{m \in \mathbb{Z}}  
             \Big[  \int_{\mathbb{R}} g (\tau) \, e^{-2 i \pi m \tau} \, d
             \tau  \Big] \, \, e^{2 i \pi m x}. \label{eq:fourier}
\end{equation}
This amounts to converting the sum in the l.h.s. for $g(x)$ into a sum over
its Fourier transformation $\tilde{g}(\tau)$ in the r.h.s. Here $g(x)$ is a
continuous function of $x$ which decreases rapidly for $x \rightarrow
\infty$.

In our case (\ref{eq:eisenstein}), using the quadratic form defined in
(\ref{eq:newq}), we can write
\begin{eqnarray}
  \sum_{m,n \neq (0,0)} \frac{1}{(m^2 +  m n i (1-T) + n^2
  T)^{s}}  = & & \\ 
  \nonumber 
 & & \hspace{-60pt}  \sum_{(m,n) \neq (0,0)} \frac{1}{(m +
   n(z+i\alpha))^{s} \, ( m + n(z- i\alpha))^s}  \label{eq:final-sum}
\end{eqnarray}
where we have defined 
\begin{equation}
  z =  \frac{i}{2} \,\,(1-T)  \quad 
       \mbox{and} \quad \alpha = \frac{1}{2} (T+1) \label{eq:zeta-alpha}
\end{equation}
we also can split the sum as $\sum_{(m,n) \neq (0,0)} = ( 2
\sum_{n=0,\,m > 0}  +2 \sum_{n>0, m \in \mathbb{Z}})$. 

\subsection{The Fourier Transformation.}
\markright{\thesection. The Fourier  Transformation}

After some algebra, which we will just sketch here, one can find the Fourier
transformation (\ref{eq:fourier}), the expression
\begin{equation}
  \label{four-sum}
  \sum_m g (x+m) = \sum_m
\,(m + x + i y +i \delta)^{-s}\, (m + x + i y -i \delta)^{-s} 
\end{equation}
leads, to
\begin{equation}
  \sum_m  g (x+m) = \sum_r \left( \int_{\mathbb{R}}\frac{1}{( w + i y +
     i \delta)^{s}\, ( w + i y - i \delta)^{s}} \,\, e^{-2 i \pi rw} dw
      \right) \, \, e^{2 i \pi r x} \,.
\end{equation}
Using the integral \cite{table}
\begin{equation}
  \label{bessel}
  \int_{-\infty}^{\infty} e^{-i v u} \, (u^2 +1)^{-s} \, du \,\, =\,\,
      \left\{ 
      \begin{array}{lcr}
        \frac{1}{\Gamma(s)}\,\, 2 \pi^{1/2} \left( \frac{|v|}{2} 
                \right)^{s-\frac{1}{2}} K_{s-\frac{1}{2}} (|v|) &
                \mbox{for} & v \neq 0 \\[5pt]
        \frac{1}{\Gamma(s)} \,\, \pi^{1/2} \Gamma(s-\frac{1}{2}) &
        \mbox{for} &   v = 0 
      \end{array} \right.
\end{equation}
where $ K_{a} (b) = \int_0^\infty \, e^{-\beta \cosh (b)} \, \cosh(a\beta)
\, d\beta$ is the modified Bessel function\footnote{The absolute value here
  only means that the sign of the argument of the Bessel function is set so
  that integral is convergent.} and, using the notation $\sigma = x+iy$ in
(\ref{four-sum}), one obtains 
\begin{eqnarray}
  \sum_m  g (x+m) & = & 2
     \left[  \right.
    \frac{1}{\Gamma(s)}\,\,  \pi^{1/2} \,\,
    \Gamma(s-\frac{1}{2} \,)\,\,
    \delta^{-2s+1} \phantom{space} \nonumber \\[5pt]  
     & & + \frac{2}{\Gamma(s)}\,\  \pi^s\,\,
       \sum_{r \neq 0}\,\,e^{2 i \pi r \sigma} 
     \,\,\delta^{-s+\frac{1}{2}} \,\, 
    |r|^{s-\frac{1}{2}} \,\, 
    K_{s-\frac{1}{2}}(2 \pi  |r| \delta) \left. \right]. 
\end{eqnarray}

The last expression can now be used to calculate the sum
(\ref{eq:final-sum}) with $\sigma = n z$ and $\delta = n \alpha$. Splitting
the sum as announced before, one obtains for (\ref{eq:eisenstein})
\begin{eqnarray}
\hspace{-1cm} \sum_{m,n \neq 0} (\pi \mathbf{Q}(T,m,n))^{-s} & = &
          2\,\pi^{-s}\, \zeta(2s) + \, \frac{2\pi^{\frac{1}{2}-s} 
          \Gamma(s-1/2)}{\Gamma(s)} \,\,  \alpha^{1-2s} \, \zeta(2s-1)
        \nonumber \\[6pt]
    & & \hspace{-1.9cm} + \,\, \frac{4}{\Gamma(s)} \,\, 
           \alpha^{1/2-s}
       \sum_{n>0,\, r \neq 0} e^{2 \pi irnz} n^{1/2-s} |r|^{s-1/2} 
           K_{s-1/2}(2\pi|r|n\alpha).
\end{eqnarray}
in this formula $\zeta(s)$ denotes the ordinary Riemann $\zeta$-function.

From previous sections we learned that the function
\begin{equation}
 2 \, G^*(s) \,\, \equiv \,\, \Gamma(s) \sum_{(m,n) \neq (0,0)}(\pi
   \mathbf{Q}(m,n,z))^{-s} \label{eq:meromor}
\end{equation}
extends to a meromorphic function in the complex plane. We also learned
that it has a pole at $s=0$ and that, once the pole is subtracted, the
resulting function is regular in the limit. We know too that
(\ref{eq:kron})
\begin{equation}
 \lim_{s \rightarrow 0} \frac{d}{ds} \,\, \zeta(s)
                       =  \lim_{s \rightarrow 0} \left( 2 G^*(s) +
                         \frac{1}{s} \right)
\end{equation}
where $\zeta(s)$ now denotes the  $\zeta$-function defined in
(\ref{eq:eisenstein}). 

Collecting all our findings so far, we can write
\begin{eqnarray}
  2 \, G^*(s) & =& \pi^{-s}\,\, \Gamma(s)\,\, \zeta(2s)\,\,
                  +   \,\,
                  \pi^{1/2 -s}\,\, \Gamma(s-1/2)\,\, \alpha^{1-2s} 
                   \zeta(2s-1) \nonumber \\[5pt]
              & & + \,\,2 \alpha^{1/2-s}\,
                   \sum_{n>0,r \neq 0} 
                  e^{2 i \pi r n z} n^{1/2 -s} |r|^{s-1/2} 
                  K_{s-1/2}(2 \pi|r| n \alpha). \label{eq:holom}
\end{eqnarray}

The function $\zeta^*(a)\pi^{-a/2}\,\, \Gamma(a/2)\,\, \zeta(a)$ is
holomorphic in the complex plane \cite{number}, it has a simple pole at
$s=0$ with residue $-1$.

Using the identity 
\begin{equation}
  K_{1/2} (x) = \sqrt{\pi/2x}\,\, e^{-x}
\label{eq:bessel}
\end{equation}
we find
\begin{eqnarray}
\label{eq:ya-casi}
  \lim_{s \rightarrow 0} \,(\,\,2\, G^*(s)\, +\,\frac{1}{s}\,) & = &
               C_0 + \,\frac{1}{6}\,\, \pi \alpha +  \nonumber \\ 
     & & \mbox{} \,\,\sum_{n,r>0} \,\, \frac{1}{r}\,\,
         (e^{2i\pi rn(z+i\alpha)} + e^{-2i\pi rn(z-i\alpha)}),
\end{eqnarray}
where $C_0$ is a constant of no importance to us and we have make use of
the value $\zeta(-1) = -1/12$.  The sums implied here can be performed with
the help of the  identity \cite{elliptic},
\begin{equation}
  \ln \,\, \eta (z) \, = \, \frac{i\pi z}{12} - \sum_{m,k>0} \,
                          \frac{1}{k} \,\, e^{2 i \pi k m z};
\end{equation}

\noindent where $\eta(z)$ is the Dedekind function. Thus we end  with the
expression
\begin{equation}
\label{eq:first-term}
   \lim_{s \rightarrow 0} \,(\,\,2\, G^*(s)\, +\,\frac{1}{s}\,) = 
                           \,\, \ln\,\eta (z+ i\alpha) +
                           \ln \, \eta(-z+i\alpha) \,\, + C_0 \\[5pt]
\end{equation}
This equation resembles the First Kronecker Limit Formula of the theory of
elliptic functions \cite{elliptic,number}.  Both terms are convergent. 

\subsection{Putting It All Together Again.}
\markright{\thesection. Putting It All Together Again.}

We recall here that we need to add the complex conjugate to the formula
(\ref{eq:first-term}), corresponding to the second term of the
$\zeta$-function expansion (\ref{eq:myzeta}) to obtain the automorphic
function. A third term should be added as well, provided in the equation
(\ref{eq:automorphic}), with the K\"ahler potential written as in
(\ref{eq:newq}) in the transformed variables $T$.

In short one finds, with the values of $z$ and $\alpha$ given by
(\ref{eq:zeta-alpha}), that the automorphic function of
$\frac{SU(1,1)}{U(1)}$ can be written as
\begin{equation}
  \label{eq:newautomorphic}
    e^{\mathscr{G}}  =   |\Delta|^2 \, e^{K} \, =\,[\,
    |\,\eta(T)\, |^4\, (T -\bar{T}) \,] \,\,+ C
\end{equation}
where we have included a factor of $i$ into the variable $T$ and the
constant term $C$ is moduli independent.  What we have obtained here is the
classical result of reference \cite{ferr91} (up to constant
moduli-independent factors).

The automorphic function is invariant under $T \rightarrow T+1$ and $ T
\rightarrow -\frac{1}{T}$, therefore it is $SL(2,\mathbb{Z})$ invariant (up
to roots of unity).

The question now is why go through all this lengthy process to find what we
already knew?  The reason for this is that the procedure described here can
be generalized to cosets beyond $\frac{SU(1,1)}{U(1)}$ and to any
$\frac{SU(1,n)}{U(1) \times SU(n)}$. We will describe the next case in the
following section.

\section{The  $SU(1,2)/ U(1) \times SU(2) $ Coset and the
  General Ansatz.}  

\markright{\thesection. The $SU(1,2)/ U(1) \times  SU(2)$ Coset...}
\label{section:su2}

The form of the automorphic function in this case, taken from
(\ref{eq:limit}), is going to depend on
\begin{equation}
  (p_i x^i + q^i \mathcal{F}_i) = p_0 x_0 + p_1 x_1 + i q_1 x_1
                                  +p_2 x_2 +i q_2 x_2.
\end{equation}
As before the integers are subject to the $SU(1,n,\mathbb{Z})$-invariant
constraint
\begin{equation}
  (p_0)^2 = (p_1)^2 + (q_1)^2+ (p_2)^2 + (q_2)^2 \label{eq:new-const}
\end{equation}
where we have also set the to zero the partner of the imaginary part of
$x_0$. 

The constraint (\ref{eq:new-const}) is also possible to solve
\cite{dickson}, in the same spirit  of the solutions provided for the
previous example. Indeed, one can write
\begin{eqnarray}
  p_0 &=& (n_1)^2 + (n_2)^2+ (n_3)^2 + (n_4)^2  \nonumber  \\ 
  p_1 &=& -(n_1)^2 + (n_2)^2+ (n_3)^2 + (n_4)^2  \nonumber \\
  q_1 &=& 2 n_1 n_2  \nonumber \\
  q_2 &=& 2 n_1 n_3  \nonumber \\
  p_2 &=& 2 n_1 n_4 \label{eq:sqrt-sol}
\end{eqnarray}
for $\vec{n} \equiv ( n_1, n_2, n_3, n_4) \in \mathbb{Z}^4$. This provides
a general solution to (\ref{eq:new-const}). However, there is one set of
numbers that are not produced by this array. Nevertheless, such solution
can be constructed using it.

\begin{table}
\vspace{5mm}
\centering
\begin{tabular}{cccccc} \hline
Solution & $p_0 $ & $p_1 $ & $q_1 $ & $p_2 $ & $q_2 $ \\  \hline
I   & \hspace{2mm} $0 \pmod{4}$ \hspace{2mm} & e & e & e & e \\
II  & \hspace{2mm} $1 \pmod{4}$ \hspace{2mm} & o & e & e & e \\
III & \hspace{2mm} $0 \pmod{4}$ \hspace{2mm} & o & o & o & o \\ \hline
\end{tabular}
\caption{ \label{table:squares}
 {\small Solutions to the constraint (\ref{eq:new-const})
    according to whether $p_i, q_i$ are \textbf{e}ven or \textbf{o}dd.}}
\end{table}

To be more precise, the square of an integer is always $0$ or $1$ modulo
$4$. That is, the l.h.s. of (\ref{eq:new-const}) equals $ 0,1 \pmod{4}$.
Since the sum in the r.h.s is a sum of four squares we can have that the
r.h.s can be \textit{any} integer. (In fact, it was Lagrange who proved
that any integer can be written as the sum of four squares.)  Since there
is the equality in the middle, only some combinations of numbers can
satisfy the condition.  The correct combinations are given in Table
\ref{table:squares}.

\subsection{First Contribution to the Automorphic Function.}
\markright{\thesection. First Contribution to the Automorphic...}

It is not difficult to see that solutions I and II can be generated by
(\ref{eq:sqrt-sol}). Solution III is generated as follows: take all numbers
in $\vec{n}$ to be odd, this means that all the expressions in the l.h.s
are even. We now divide by two all the expressions of the r.h.s of
(\ref{eq:sqrt-sol}). The resulting solution corresponds to case III.  We
will concentrate first in the contribution of the cases I and II, the third
will be discussed at the end.

Implementing the solution given above leads to the $\zeta$-function
\begin{eqnarray}
  \label{eq:riemann-too}
 \mathbf{\zeta} (s)& = & \frac{1}{\Gamma(s)}\,\, \int_0^{\infty} d \tau \, 
       \tau^{s-1} \sum_{\vec{n} \in \mathbb{Z}^4} \Big(  \exp [ -\tau \pi
         \,\mathcal{Q}(\vec{n},t,A)] -1 \Big)  \nonumber \\[5pt]
  & = & \frac{1}{\Gamma(s)}\,\, \int_0^{\infty} d \tau \, 
       \tau^{s-1}  (\mathbf{\Theta}(\tau,t,A)  -1)
\end{eqnarray}
\noindent
where we have the quadratic form in terms of the inhomogenous (physical)
coordinates $x_0 =1, \,\, x_1 =t, \, \, x_2=A$. The quadratic form
$\mathcal{Q}$ is given by
\begin{eqnarray}
\label{eq:omega}
  \mathcal{Q}(\vec{n},t,A)\,\,& = &  \,\, \vec{n}^T \, \cdot \Omega_2
  \cdot \vec{n}   \nonumber \\[8pt]
 \mbox{with}  \qquad \qquad \qquad \Omega_2 \,\,& = &
 \,\, \left(  
    \begin{array}{cccc}
      1-t & it & i A & A \\
      it & 1+t & 0 & 0 \\
      iA & 0 & 1+t & 0 \\
      A & 0 & 0 & 1+t
    \end{array} \right);
\end{eqnarray}
which satisfies $\det \Omega_2 = (1+t)^2$, i.e. it is invertible. (Using a
solution with $q_0 \neq 0$ leads to a non-invertible matrix.) 

The resulting Mellin transformation is not easy to analyze in this case.
However, we can still work out some form of the Fourier transformation in
the other half of the procedure. A general form of such expansion is
provided by the recursive use of the Poisson summation formula
(\ref{eq:fourier}) as shown in reference \cite{eliz97}. If
$\zeta_{\Omega_{2}}(s) = \sum_{0 \neq n \in \mathbb{Z}^d}
(\mathcal{Q})^{-s} $ this procedure, in general, renders the
$\zeta$-function expansion
\begin{eqnarray}
  \label{eq:elizalde}
  \zeta_{\Omega_2} (s)& =& 2 a^{-s} \zeta(s) + \sqrt{\frac{\pi}{a}}\,
  \frac{\Gamma(s-1/2)}{\Gamma(s)} \,\, \zeta_{\Omega_{\mathrm{Red}}}
  (s-1/2) \, + \nonumber \\[5pt]
 & & \frac{4 \pi^s}{a^{s/2+1/4}\,\Gamma(s)} \, \sum_{\vec{0} \neq \vec{n}'
   \in \mathbb{Z}^{d-1}} \, \sum_{m=1}^{\infty} \,\left[ \right. \cos
  (\frac{\pi}{a} m \vec{b} \cdot \vec{n}') \, m^{s-1/2} (\vec{n}' \cdot 
    \Omega_{\mathrm{Red}} \cdot \vec{n}')^{1/4-s/2} \nonumber \\
  & & \hspace{5.3cm} \times K_{s-1/2}(\frac{2\pi}{\sqrt{a}} m \sqrt{
    \vec{n}' \cdot \Omega_{\mathrm{Red}} \cdot \vec{n}'})  \left. \right];
\end{eqnarray}
where $d$ is the dimensionality of $\Omega_2$, $\,\,a =\Omega_{2_{00}}$,
$\vec{b} = (\Omega_{2_{2,1}},\Omega_{2_{2,2}}, \ldots \Omega_{2_{2,d}})$
and $$\Omega_{\mathrm{Red}}~=~\Omega_{2_{d-1 \times d-1}} - (1/4a)
\,\vec{b} \otimes \vec{b},$$ with $\Omega_{2_{d-1 \times d-1}}$ being the
reduced matrix $\Omega_2$ without the first column and row. 

This is a recursive formula which leads to an analytic --although
complicated-- Fourier expansion of the automorphic function. One can,
however, use the fact that $\Omega_2$ has some nice symmetries to obtain
the final formula. Before this, it is better to write the quadratic form as
we did in (\ref{eq:newq}). In this case we change  variables as follows
\begin{equation}
 t \rightarrow  T= \frac{1-t}{1+t} \qquad A \rightarrow A' = \frac{\sqrt{2}
   A}{1+t}.
\end{equation}
Under which the quadratic $\mathcal{Q}(n,t,A) \rightarrow
\mathcal{Q}(n,T,A')$ has the matrix
\begin{equation}
  \Omega'_2 = \left( 
    \begin{array}{cccc}
      T & \frac{1}{2} i(1-T) &  \frac{1}{\sqrt{2}}i A' & \frac{1}{\sqrt{2}}
      A'\\ 
      \frac{1}{2} i(1-T) & 1 & 0 & 0 \\
      \frac{1}{\sqrt{2}}i A' & 0 & 1 & 0 \\
      \frac{1}{\sqrt{2}} A' & 0 & 0 & 1
    \end{array} \right) \\[6pt]
\end{equation}
and the K\"ahler potential transforms as 
\begin{equation}
  \label{eq:trans-prep}
  1 - t \bar{t} -A \bar{A}
\rightarrow 2( T + \bar{T} - A' \bar{A'}).
\end{equation}
Correspondingly, the
$\zeta$-function is given by 
\begin{eqnarray}
  \zeta_{\Omega'_2} (s) & = \! & \frac{1}{\Gamma(s)}\,\,\int_0^{\infty} d
         \tau \, \tau^{s-1} \sum_{\vec{0} \neq \vec{n} \in \mathbb{Z}^4}
         \exp [ -\tau  \pi \,\mathcal{Q}(\vec{n},T,A')]  \\[6pt]
   &= &  \! \frac{1}{\Gamma(s)}\,\, \int_0^{\infty} d
         \tau \, \tau^{s-1} \sum_{\vec{0} \neq \vec{n} \in \mathbb{Z}^4}
         e^{-\tau  \pi \, [T (n_1)^2 + \sum_{j=2}^{4} (n_j)^2 
         + n_1 n_2 i (1-T) + n_1 n_3 i \sqrt{2} A' +n_1 n_4 \sqrt{2} A']}
       \, . \nonumber 
\end{eqnarray}

Completing squares for each $n_j$ one obtains
\begin{equation}
   \zeta_{\Omega'_2} (s) = \frac{1}{\Gamma(s)}\, \int_0^{\infty}\! 
         d\tau \, \tau^{s-1}\hspace{-3mm} \sum_{ \vec{0}
           \neq            \vec{n} \in \mathbb{Z}^4}  \hspace{-2mm}
          e^{-\tau  \pi \, [\frac{1}{4} (1+T)^2  (n_1)^2 + (n_2 + \frac{i
              n_1}{2} (1-T))^2 + (n_3 + \frac{i n_{1}}{\sqrt{2}} A')^2 + 
            (n_4 + \frac{n_{1}}{\sqrt{2}} A')^2]} \, ;
\end{equation}
where we can split the sum into $ \sum_{\vec{0} \neq \vec{n} \in
  \mathbb{Z}^4} = \sum_{\vec{0} \neq \vec{N} \in \mathbb{Z}^3 , n_1 = 0} +
\sum_{n_1 \neq 0, \vec{N} \in \mathbb{Z}^3} \equiv \zeta_I + \zeta_{II} $,
and define $\vec{N} \equiv (n_2, n_3, n_4)$.  The second term in this sum
can be analyzed using the properties of Mellin transformations, for that
term we obtain
\begin{equation}
 \zeta_{II} =   \frac{1}{\Gamma(s)} \,\, \int_0^{\infty} d \tau \,
       \tau^{s-1} \sum_{ \vec{N}\neq \vec{0} } 
      e^{- \pi\tau [(n_2)^2 + (n_3)^2 + (n_4)^2]}
\end{equation}
which can be written
\begin{equation}
 \mathbf{M} \,[ \phi(s)] =   \int_0^{\infty} d \tau \,\tau^{s-1}  \phi
  (\tau) 
\end{equation}
with 
\begin{equation}
  \phi (\tau) = [\theta_3(\tau)]^3 - 1 \quad \mbox{and} \quad
  \theta_3(\tau) = \sum_n     e^{-\pi \tau n^2} \, . 
\end{equation}
Jacobi's $\theta_3(\tau)$ function satisfies $ [\theta_3(\tau)]^3 =
\tau^{-3/2} \, [\theta_3(\tau^{-1})]^3$.  Therefore, we have the functional
equation
\begin{equation}
  \phi (\tau^{-1}) = \tau^{3/2} \phi (\tau) + \tau^{3/2} - 1.
\end{equation}
Thus we have simple poles at $s=0$ and $s=3/2$ with residues $-1$ and $+1$
respectively. 

The first part of the sum is more involved. First we make use of the
Poisson formula \cite{lang},
\begin{equation}
  \sum_{n \in \mathbb{Z}} e^{-\pi  \tau (n+xm)^2} = \frac{1}{\sqrt{\tau}}
      \,\, \sum_{n \in \mathbb{Z}} e^{2\pi i xmn} e^{-\pi n^2/\tau}
\end{equation}
over every component of $\vec{N}$. Then we can integrate with the help of
the identity 
\begin{equation}
  \int_0^\infty dx \,x^{\nu-1}\, e^{-\frac{\beta}{x} -\alpha x } = 2\,
    \Big(\frac{\beta}{\alpha}\Big)^{\nu/2}\,  K_{\nu}(2\sqrt{\beta
      \alpha}), 
\end{equation}
for $\beta,\alpha \neq 0$. 

Doing all this, we get 
\begin{eqnarray}
   \zeta_{I} & = &
   \frac{1}{\Gamma(s)}\,\, \Big[ \sum_{n_1 \neq 0, \vec{N}\in
      \mathbb{Z}^3 } e^{i \pi n_1[n_2 i (1-T) + n_3 \sqrt{2} i A'
                   + n_4 \sqrt{2} A']} \nonumber \\[5pt] 
      & &  \qquad \times  \int_0^{\infty} d \tau \, \tau^{(s-3/2)-1} 
           e^{-(n_1)^2 \pi \tau (\frac{1+T}{2})^2 -\frac{\pi}{\tau} (
            (n_2)^2 + (n_3)^2 + (n_4)^2)} \,\, \Big]. 
          \label{eq:ya-casito}
\end{eqnarray}

Considering the cases $\vec{N}= \vec{0}$ and $\vec{N} \neq \vec{0} $
separately (and integrating the second part over $\tau$)  we obtain
\begin{eqnarray}
   \zeta_{\Omega'_2} (s) & = & \zeta_{II} \,\, +  \,\,
      \left(\frac{1+T}{2}\right)^{-2s+3} \,\,\frac{1}{\Gamma(s)}\,\,
     \int_0^\infty d \tau  \tau^{(s-3/2)-1}   [\theta_3(\tau) -1]  \,\, +
     \nonumber \\[5pt] 
  & & \hspace{-10mm} + \,\,  \frac{1}{\Gamma(s)}   \,\,
     \sum_{0 \neq   n_1 \in \mathbb{Z}, \vec{N}  \neq \vec{0}} 
        \Big[ e^{i \pi n_1 [ n_2 i (1-T) + n_3 \sqrt{2} i A'
                   + n_4 \sqrt{2} A']}  \\[5pt]
 & & \qquad \times \,\, 2^{s-1/2} \, |\vec{N}|^{s-3/2} \, n_1^{-s+3/2}\,
             (1+T)^{-s+3/2} K_{s-3/2} \, (\pi n_1 |\vec{N}| (1+T)) \Big]
            \,\, ; \nonumber \label{eq:complique}
\end{eqnarray}
we have to notice here that this expression can not be brought to a form
similar to the case of $SU(1,1)/U(1)$, where the indexes and exponentials
were $1/2$ instead of $3/2$. That is, the presence of the other moduli
cannot be switched off simply by setting the moduli and the winding and
momentum numbers to zero. If one is to consider that case, the
approximation should be made in the previous step of the calculation and
not implemented as truncations in the Fourier expansion
(\ref{eq:complique}).

The pole structure of the above formula is not difficult to study. However,
according to what we have already learned, we are interested in the function
($\Gamma(s) \zeta_{\Omega'_2} (s)$) which extends to a meromorphic function
in the complex plane, except for poles at certain values of $s$.

Since the third term is convergent as long as the exponentials are in the
upper half plane, the poles are given by the first two terms. We find that
--after the use of the properties of the Mellin transformation-- the
function $2 \mathbf{G}(s) \equiv \Gamma(s) \pi^{-s} \sum_{n \in
  \mathbb{Z}^4} (\mathcal{Q}(n,t,A'))^{-s}$ (defined as in
\ref{eq:meromor}) has simple poles located at $s=0$ and $s=2$ with residues
$-1$ and $+1$.  Hence, after making a Laurent expansion and subtracting
the pole, we can proceed to study the limit $s \rightarrow 0$ in the
expression (\ref{eq:complique}).

From (\ref{eq:complique}) we therefore obtain
\begin{eqnarray}
\hspace{-1.3cm} \lim_{s \rightarrow 0}  (2 \mathbf{G}(s) + 
  \frac{1}{s}) & = & \,\, C_0   \,\, + \,\, \frac{\pi^2}{45}\,
   \left(\frac{1+T}{2} \right)^3 \, +  \nonumber \\[5pt]      
   & & \hspace{-7mm}  \left(\frac{1+T}{2} \right)^{3/2} \! \!\!\!
    \sum_{0 \neq n_1 \in \mathbb{Z}, \vec{N}  \neq \vec{0}}  \Big[ e^{i
        \pi n_1 [n_2 i (1-T) + n_3 \sqrt{2} i A'+ n_4 \sqrt{2} A']}
      \nonumber \\[5pt]  
  & & \qquad \qquad \qquad \qquad\,\,\,\, \times \,  |\vec{N}|^{-3/2}
      n_1^{3/2}   K_{-3/2} (\pi n_1   |\vec{N}| (1+T))  \Big] 
 \label{eq:mascompl}
\end{eqnarray}
where $ C_0$ is a $T$-independent constant. The recurrence relations
\cite{table} for the Bessel function $K_{3/2}(z) = K_{-3/2} (z)$ allow us
to write $K_{3/2}(z) = z^{-1} \, K_{1/2}(z) + K_{1/2}(z)$, where we
can also make use of (\ref{eq:bessel}). So, we can write
\begin{equation}
  K_{3/2}(\pi n_1 |\vec{N}| (1+T)) = \frac{1 + \pi n_1  |\vec{N}|
    (1+T)}{\pi n_1  |\vec{N}| (1+T)} \, (2\pi  n_1
  |\vec{N}|(1+T)))^\frac{-1}{2 }
          \, e^{-\pi n_1  |\vec{N}| (1+T)}. \\[5pt]
\end{equation}

\noindent Introducing this expression into (\ref{eq:mascompl}), we find
\begin{eqnarray}
  \hspace{-1.3cm}  \lim_{s \rightarrow 0}  (2 \mathbf{G}(s) + \frac{1}{s})
  & = & \,\, C_0 
     \,\, + \,\, \frac{\pi^2}{45}\,   \left(\frac{1+T}{2} \right)^3 \,\, + 
     \nonumber \\[5pt] 
 & & \hspace{-7mm} \frac{1}{4} \sum_{0 \neq n_1 \in \mathbb{Z}, \vec{N}
   \neq \vec{0}} 
 \Big[e^{i  \pi n_1 [n_2 i (1-T) + n_3 \sqrt{2} i A'+ n_4 \sqrt{2} A' + i
         |\vec{N}| (1+T)]}  \nonumber \\[5pt]
 & & \qquad \qquad \qquad \qquad\,\,\,\, \times \,\Big(\pi |\vec{N}|^{-3} +
          |\vec{N}|^{-2} n_1 (1+T)\Big) \Big]. \label{eq:casi-final}
\end{eqnarray}
This is essentially the last formula of this section. Presumably there
could be a way of writing the sums in (\ref{eq:casi-final}) in terms of
known functions but that looks like a difficult task.

Summarizing, we have that $\Psi(\Omega_2(T,A')) = \lim_{s \rightarrow 0} (2
\mathbf{G}(s) + \frac{1}{s})$ and therefore the automorphic
function (\ref{eq:automorphic}), considering only the contribution
from the solutions I and II of table \ref{table:squares}, is given by
$e^{\mathscr{G}} = e^{4 \, \mathrm{Re} \Psi(\Omega_2(T,A'))} \, |T+\bar{T}
+ A' \bar{A'}|$.

A general Ansatz for finding the automorphic functions  of coset
manifolds of the form $SU(1,n) / SU(n) \times U(1)$ can be deduced from the
above discussion. Indeed, the recursive use of the formulas
(\ref{eq:sqrt-sol}) is not difficult to implement and the rest of the
equations are also easy to generalize. The inclusion of more moduli $A_i$
into our formulas would only increase the dimension of the matrix
$\Omega_2$ but its general structure remains the same
\begin{equation}
  \mathbf{\Omega'}_n = \left( 
    \begin{array}{ccccccc}
      T & \frac{1}{2} i(1-T) &  \frac{1}{\sqrt{2}}i A' & \frac{1}{\sqrt{2}}
      A' & \cdots &  \frac{1}{\sqrt{2}}i A'_n & \frac{1}{\sqrt{2}}
      A'_n \\ 
      \frac{1}{2} i(1-T)     & 1 & 0 &   & \cdots & & 0 \\
      \frac{1}{\sqrt{2}}i A' & 0 & 1 & 0 & \cdots  & & 0 \\
      \frac{1}{\sqrt{2}} A'  & 0 & 0 & 1 & 0 & \cdots & 0 \\
       \vdots &  \vdots &  & \cdots  &   & & \vdots \\
      \frac{1}{\sqrt{2}} A'_n & 0 & & \cdots &  & & 1 
    \end{array} \right). \\[6pt]
\end{equation}

So the $\zeta$-function function to be calculated is the Mellin
transformation
\begin{equation}
  \zeta_{\mathrm{gen}}(s) = \frac{1}{\Gamma(s)} \,\, \int_0^\infty d\tau
  \,\,\tau^{s-1} \,\,[ \Theta_{\mathbf{\Omega'}_n} (\tau) -1]. 
  \label{eq:genzeta} 
\end{equation}

It is also easy to see that the positions of the poles of the Fourier
expansion of $\Gamma(s)\zeta_{\mathrm{gen}}(s)$ are affected; nonetheless,
one can see that there is always a pole at the position $s=0$ (due to the
particular points $n_1 \neq 0, \, \vec{N} =0$ for example).  Once this pole
is subtracted we obtain a regular expression for the automorphic
function.  The final form for a general coset will depend on the
recursion formulas for the Bessel functions, but the structure of the
solution is similar in form to (\ref{eq:ya-casito}).

\subsection{Second Contribution to the Automorphic Super\-potential and
  Final Expression.}
\markright{\thesection. Second Contribution to the...}

We saw in the previous section that the $SU(1,n)$-orbits described by the
constraint (\ref{eq:new-const}) can be divided in three types given in
table \ref{table:squares}.  However, we have only considered the solutions
provided by the cases I and II. We will now see what happens to the
extra piece, coming from the solution III. In short we just find that the
Riemann theta function (\ref{eq:riemann-too}) gets extra terms and gets
promoted to a $\theta$-function whose general expression is given by the
definition (\ref{eq:character}). 

The solution to constraint, following the procedure explained after the
equation (\ref{eq:sqrt-sol}), is given by
\begin{eqnarray}
  p_0 &=&  2\, [\phantom{-}(n_1)^2 + (n_2)^2+ (n_3)^2 + (n_4)^2] + 
           2 \, [\phantom{-} n_1 + n_2+ n_3 + n_4] +2  \nonumber  \\ 
  p_1 &=& 2 \, [ -(n_1)^2 + (n_2)^2+ (n_3)^2 + (n_4)^2]
         +2 \, [-n_1 + n_2+ n_3 + n_4] +1 \nonumber  \\ 
  q_1 &=& 4 \, n_1 n_2 + 2 (n_1 + n_2 ) + 1 \nonumber  \\ 
  q_2 &=& 4 \, n_1 n_4 + 2 (n_1 + n_4 ) + 1 \nonumber  \\ 
  p_2 &=& 4 \, n_1 n_3 + 2 (n_1 + n_3 ) + 1 
\end{eqnarray}
following the same steps as before, we find that the  theta function
involved is 
\begin{equation}
  \mathbf{\Theta}'
  [\Omega_2(t,A)] =  \sum_{\vec{n} \in \mathbb{Z}^4} e^{-2 \pi \tau 
    (\vec{n}
    \cdot \Omega_2 \cdot \vec{n} + 2 \vec{n} \cdot \vec{\mathbf{Z}} +
    \mathbf{B})} \, , \label{eq:more-riem}
\end{equation}
where $\Omega_2$ is the same as in (\ref{eq:omega}). 

Here the product ``$\cdot$'' is taken with an Euclidean metric and we have
introduced
\begin{equation}
  \vec{\mathbf{Z}} \equiv \left[ 
    \begin{array}{c}
      1 - (1-i) t + (1+i) A \\
      1 + (1+i) t \\
      1 + t + i A\\
      1 + t + A
    \end{array} \right] \qquad \mbox{and} \qquad 
  \mathbf{B} \equiv 2 + (1+i) (t +A). \label{eq:zyb}
\end{equation}

The $\zeta$-function, in the lines seen many times before here can be
expanded in a Fourier series. This is possible because of the properties
(\ref{eq:theta-inv}) and (\ref{eq:theta-inv1}).  This Fourier expansion of
(\ref{eq:more-riem}) can be found using the results from reference
\cite{eliz97}. There they use of the Poisson summation (\ref{eq:poisson}),
generalized as follows
\begin{equation}
 \sum_{\vec{n} \in \mathbb{Z}^d}\, \exp \, \Big( -\frac{1}{2}\, \vec{n}
    \cdot \mathcal{A} \cdot \vec{n} + \vec{b} \cdot \vec{n} \Big) =
    \frac{(2 \pi)^{d/2}}{\sqrt{\det \mathcal{A}}} \,\, \sum_{\vec{m} 
     \in \mathbb{Z}^d} \, \exp \, \Big( \frac{1}{2} (\vec{b} + 2 i\pi
   \vec{m})\cdot \mathcal{A}^{-1} \cdot (\vec{b} + 2 i\pi \vec{m}) \Big)
\label{eq:more-poisson}
\end{equation}
where $d$ is the dimensionality of the complex-valued matrix
$\mathcal{A}$. 

The $\zeta$-function $\sum_{\vec{n}} [\mathcal{Q} (\vec{n},t,A)]^{-s}$ that
we defined above is promoted in this case, by the means of a Poisson
summation, to $\sum_{\vec{n}} [\mathcal{Q} (\vec{n}+ \vec{c},t,A) +
\mathbf{B}]^{-s}$ where $\vec{c} \in \mathbb{C}^d$ and $\mathbf{B}$ is a
term which does not contribute to the sum as in (\ref{eq:zyb}). Formally
one can use (\ref{eq:more-poisson}) to obtain, using the formula provided
by Elizalde in reference \cite{eliz97}, \vspace{4pt}
\begin{eqnarray}
\hspace{-10mm} \zeta_{\mathbf{\Theta}'} (s) \,\,& =&  \,\,\,
     \frac{\mathbf{B}^{2-s} 
    \Gamma(s-2)}{\sqrt{\det \Omega_2} \Gamma(s)} \, \,+ \, \,
          \frac{2^{s/2+1} \pi^{s-2} \mathbf{B}^{-s/2+1}}{\sqrt{\det
              \Omega_2} \Gamma(s)} \nonumber \\[7pt]
  & &   \hspace{-10mm} + (2\pi)^{s/2+1}  \sum_{\vec{n}
        \in \mathbb{Z}^d_+} 
    \cos (2i \vec{n} \cdot \vec{\mathbf{Z}}) \, (\vec{n} \cdot
    \Omega_2^{-1} 
  \cdot \vec{n})^{s/2-1} \, K_{s-2} \,\Big( 2 \sqrt{\pi \mathbf{B}\vec{n}
    \cdot  \Omega_2^{-1} \cdot \vec{n}} \Big) 
\end{eqnarray}
in which $ \mathbb{Z}^d_+$ indicates that only half of the integers
contribute to the sum, for example those with positive first entry will
contribute and not those with a negative one.

The last expression is quite formal; indeed, we could have written
something similar for the case we studied rigorously in the previous
section.  It is reassuring to see the same structure that we found before
for the automorphic function emerging here as well.

The transformation rule for Jacobi's $\theta$-function 
\begin{equation}
 \label{eq:theta3}
  \sum_{n \in \mathbb{Z}} e^{\pi i n^2 T+2\pi in v} = \sqrt{\frac{T}{i}}
    \, \sum_{n \in \mathbb{Z}} e^{\pi i T (n +v)^2}
\end{equation}
allow us to repeat the same steps of the previous
calculation and perform a Fourier expansion in the same lines that we have
already tried. Hence, we split the sum --and the $\zeta$-function-- in
(\ref{eq:more-riem}) as
\begin{equation}
  \label{eq:part-sum}
 \sum_{(n_1,n_2,n_3,n_4) \neq (0,0,0,0)}  =  \sum_{n_1 = 0, \vec{N} \neq
    \vec{0}} + \sum_{n_1 \neq 0, \vec{N} \neq \vec{0}} + \sum_{n_1 \neq 0,
    \vec{N} = \vec{0}}   \, .  \\[6pt]
\end{equation}

Using the same kind of manipulation of the previous section for each of the
components of the $\zeta$-function {\boldmath $\zeta_{\Theta'}$} $(s)
\equiv $ {\boldmath $\zeta_I + \zeta_{II} + \zeta_{III}$} (labeled
according to the sums in \ref{eq:part-sum}), we obtain
\begin{eqnarray}
  \label{eq:zetaI}
 \hspace{-3cm} 
  \mbox{\boldmath $ \zeta_I$} & = & \sum_{\vec{N} \neq \vec{0}}    
   \frac{\exp[\pi i(\sum_{j=2}^4 n_j Z_j)]}{2^{s-1} \Gamma(s)} \,\, 
   |\vec{N}|^{s-3/2} \nonumber \\
  & & \,\,\,\,\,\,\, \times \,\,\,\, \Big(B - \frac{1}{4}\, 
     \sum_{j=2}^4 Z_j^2 \Big)^{-s/2 + 3/4} \, K_{s-3/2} \Big(2\pi |\vec{N}|
     \Big[B - \frac{1}{4}\,\sum_{j=2}^4 Z_j^2\Big]^{1/2} \Big)\,;
\end{eqnarray}
where we have obviously used the notation established in (\ref{eq:zyb})
duly transformed from $\mathbf{B}(t,A)$ to $B(T,A')$ and from
$\mathbf{Z}_i(t,A)$ to 
$Z_i(T,A')$.

As for the second and third parts we find

\begin{equation}
  \label{eq:zetaII}
  \mbox{\boldmath $ \zeta_{II}$}  = \!\! \sum_{n_1 \neq 0, \vec{N} \neq
  \vec{0}}  \frac{\exp[\pi i \sum_{j=2}^4 n_j(n_1 \rho_j+ Z_j)]}{2^{s-1}
  \Gamma(s)} \,\, |\vec{N}|^{s-3/2} \,\, 
                  \gamma_{n_1}^{-s/2+3/4} K_{s-3/2}
  \Big(2\pi |\vec{N}| \sqrt{\gamma_{n_1}}\Big)
\end{equation}
and 
\begin{equation}
  \label{eq:zetaIII}
    \mbox{\boldmath $ \zeta_{III}$}  = 2 \sum_{n_1 \neq 0}
           \frac{e^{\pi n i Z_1/T}}{(2T)^s \Gamma(s)} \,\, 
           |n_1|^{s-1/2} \lambda^{-s/2+1/4} \, K_{s-1/2} (2 |n| \pi
           \sqrt{\lambda}) \\[8pt] 
\end{equation}

\noindent where $\rho_{j} \equiv (\Omega_2)_{1j}$ for $j=2,3,4$ and 
\begin{eqnarray}
 \gamma_{n_1} & = & (n_1)^2 T + B - \frac{1}{4} \sum_{j=2,3,4} (n_1
 \rho_j + Z_j), \nonumber \\
 \lambda & = & \frac{ 4 T B + Z_1^2}{4 T^2} \,.
\end{eqnarray}

Here it is possible to take the limit $s \rightarrow 0$ of $\Big(
\Gamma(s)$ {\boldmath $\zeta_{\Theta'}$} $(s) \Big)$ and use the recurrence
relations for the Bessel functions to provide a final expression in terms
of $T$ and $A'$. Each of the functions $\Gamma(s)${\boldmath $\zeta_I$},
$\Gamma(s)${\boldmath $\zeta_{II}$} and $\Gamma(s)${\boldmath $
  \zeta_{III}$} is regular in that limit.

Putting together the expressions (\ref{eq:casi-final}) + (\ref{eq:zetaI}) +
(\ref{eq:zetaII}) + (\ref{eq:zetaIII}) + c.c. at $s=0$, we obtain the final
expression for the automorphic function that we are trying to find. We
obtain $e^{\mathscr{G}} = e^{4 \, \mathrm{Re} \Psi(\Omega_2(T,A'))} \,
|T+\bar{T} + A' \bar{A'}|$ with $\Psi(\Omega_2(T,A'))$ given in this case
by
\begin{eqnarray}
 \Psi(\Omega_2(T,A')) &=& \,\, C_0   \,\, + \,\,
   \frac{\pi^2}{45}\, 
   \left(\frac{1+T}{2} \right)^3 \, +  \nonumber \\[5pt]      
   & &  \left(\frac{1+T}{2} \right)^{3/2} \! \!\!\!
    \sum_{0 \neq n_1 \in \mathbb{Z}, \vec{N}  \neq \vec{0}}  
     \exp \,\Big[2 i
        \pi n_1 \Big(\sum_{j=2}^4 n_j \rho_j \Big) \Big] \nonumber \\[5pt]
  & & \qquad \qquad \qquad \qquad \qquad \qquad\,\,\,\, \times \,
      |\vec{N}|^{-3/2}  n_1^{3/2}   K_{-3/2} (\pi n_1   |\vec{N}| (1+T)) 
      \nonumber \\[5pt]
  & & + \sum_{\vec{N} \neq \vec{0}}    
   \exp \Big[\pi i\Big(\sum_{j=2}^4 n_j Z_j \Big) \Big] \, |\vec{N}|^{-3/2}
   \nonumber \\[5pt] 
  & & \qquad  \qquad \,\, \,\,\,\, \times \,
    \Big(B - \frac{1}{4}\,  
     \sum_{j=2}^4 Z_j^2 \Big)^{+ 3/4} \, K_{-3/2} \Big(2\pi |\vec{N}|
     \Big[B - \frac{1}{4}\,\sum_{j=2}^4 Z_j^2\Big]^{1/2} \Big) 
     \nonumber \\[6pt]
  & & + \!\!\!\! \sum_{n_1 \neq 0, \vec{N} \neq
  \vec{0}} \exp\Big[\pi i \sum_{j=2}^4 n_j \Big(n_1 \rho_j+ Z_j\Big)\Big] 
   \,\, |\vec{N}|^{-3/2} \,\, \gamma_{n_1}^{+3/4} K_{-3/2}
  \Big(2\pi |\vec{N}| \sqrt{\gamma_{n_1}}\Big) \nonumber \\[7pt]
  & &  +  \,\, \sum_{n_1 \neq 0} \,\, \,
           \exp \Big[\pi |n_1| i Z_1/T \Big]   \,\, 
           |n_1|^{-1/2}\, \lambda^{+1/4} \, K_{-1/2} (2 |n_1| \pi
           \sqrt{\lambda}). \label{eq:finalmente}
\end{eqnarray}

$K_{3/2}(x)$ can be expressed in terms of the exponential functions and it
is also possible to find an asymptotic form for this formula, for both the
$x \rightarrow 0$ and the $x \gg 15/8$ limits. For instance, taking the
first of these two limits renders
\begin{eqnarray}
 \Psi(\Omega_2(T,A'))_{\mathrm{asymp}} &=& \,\, C_0   \,\, + \,\,
    \frac{\pi^2}{45}\, 
    \left(\frac{1+T}{2} \right)^3 \, + \,  \frac{1}{2} \sum_{n\neq 0}
    \frac{1}{|n|}\, e^{\pi i n Z_1/T + 
        2 \pi i n \sqrt{\lambda}}  \nonumber \\[5pt]
  & &  \left(\frac{1+T}{2} \right) \! \!\!\!
    \sum_{0 \neq n_1 \in \mathbb{Z}, \vec{N}  \neq \vec{0}}  
     \frac{n_1}{|\vec{N}|^2} e^{2 \pi i n_1\vec{N}\cdot \vec{\rho} +
       \pi i |\vec{N}| (1+T)}  \, +\nonumber \\[5pt]
  & & \frac{(4B - |\vec{Z}|^2)^{1/2}}{4} \,\, \sum_{\vec{N}  \neq \vec{0}}
        \frac{1}{|\vec{N}|^2} \, e^{\pi i \vec{N}\cdot \vec{Z} + \pi i 
          |\vec{N}| \sqrt{4B - |\vec{Z}|^2}} \, +  \nonumber \\[5pt]
  & & \sum_{0 \neq n_1 \in \mathbb{Z}, \vec{N}  \neq \vec{0}}
        \frac{\sqrt{\gamma_{n_1}}}{\sqrt{2}|\vec{N}|^2} e^{\pi i n_1 
            (\vec{N}\cdot \vec{Z} + \vec{N}\cdot \vec{\rho}) +
            2 \pi i |\vec{N}| \sqrt{\gamma_n}} \; ; \label{eq:asympt}
\end{eqnarray}
where we have used the convention $\vec{\rho} = \rho_{j=2,3,4}$ and
$\vec{Z} = Z_{j=2,3,4}$. Here we can note that the truncation
$\vec{N}=(m,0,0)$ and $A'=0$ does not render the results of the previous
section, unless it is implemented \textit{before} the integrations and
re-summations are carried out. Otherwise, one is just visiting a particular
point, but not a special one, of the function $ \Psi(\Omega_2(T,A'))$.

\section{The Action of the Symplectic Group.}
\markright{\thesection. The Action of the Symplectic Group.}
\label{section:symplectic}

There is some similarity between the matrices $\Omega(T,A_i)$ of the
previous section and period matrices of compact Riemann surfaces
\cite{cand91-2,cand91,cand90}.  For instance, it has been noticed that
manifolds with Special K\"ahler geometry (like the moduli space of
Calabi-Yau manifolds) give rise to matrices which transform in the same way
as period matrices. Here we will see another example of such relation, the
profound meaning of which is yet to be clarified.

The moduli of Calabi-Yau manifolds are of two different classes, those
parametrizing the deformations of the Complex and those parametrizing the
deformations the K\"ahler structures. The dimensionality of each of these
pieces is given by the elements of the Hodge diamond which contains the
information about the $(2,1)$ and $(1,1)$-forms that parametrize the
deformations.

The holomorphic three-form $\mathbb{\Psi}$ that describes the variations to
the complex structure can be expanded in terms of a canonical homology
basis of $H_3(\mathcal{M},\mathbb{Z}),$ where $\mathcal{M}$ is the
Calabi-Yau manifold in which the compactification of the string is taking
place. If we denote this basis by $(\mathfrak{A}^a, \mathfrak{B}_b)$, with
$a,b = 0, 1 \ldots b_{21}$, and the dual cohomology basis by $(\alpha_a,
\beta^b)$, so we have the symplectic relations
\begin{equation}
  \label{eq:periods}
  \int_{\mathfrak{A}^b} \alpha_a = \int_{\mathcal{M}} \alpha_a \wedge
     \beta^b = \delta^b_a  \qquad  \int_{\mathfrak{B}^a} \beta^b =
     \int_{\mathcal{M}}  \beta^b  \wedge  \alpha_a  = -\delta^b_a .
\end{equation}
The periods of \cite{cand90} $\,\mathbb{\Psi}$ are given by $z^a \equiv
\int_{\mathfrak{A}^a} \!\mathbb{\Psi} \,$ and $\,\partial_a \mathscr{F} =
\mathscr{F}_a \equiv \int_{\mathfrak{B}^a}\! \mathbb{\Psi}$. Where we
recognize the moduli and prepotential of special K\"ahler geometry.

A symplectic transformation over the basis $(\alpha_a, \beta^b)$ leads to
the transformation rule 
\begin{equation}
  \left( 
    \begin{array}{c}
      \tilde{z}_a \\ \tilde{\mathscr{F}}_a
    \end{array} \right) =  \left(
  \begin{array}{cc}
    U & V \\ W & Z 
  \end{array} \right) \,  \left( 
    \begin{array}{c}
      {z}_a \\ {\mathscr{F}}_a
    \end{array} \right) \label{eq:vec-period}
\end{equation}
for the symplectic vector.  

The symplectic group has a natural action over the periods matrix
$\Omega\,(t, A_i)$ of the previous sections (to be precise a period matrix
of a compact Riemann surface has dimension $g \times 2g$ usually of the
form $\Pi =(\mathbb{1}, \Omega)$, they also obey the Riemann bilinear
conditions, which are equivalent to being in the Siegel upper half space
$\mathfrak{H}_g$ that we discussed earlier).  Indeed, a symplectic
transformation on the basis of $H_3(\mathcal{M},\mathbb{Z})$ , induces a
transformation similar to the transformation rule for period matrices of
compact Riemann surfaces, i.e.
\begin{equation}
  \tilde{\Omega} = ( U \Omega + V) \, ( W \Omega +Z)^{-1}.  
 \label{eq:comp-riem} 
\end{equation}
This is also very similar to the transformation rules that one finds in
toroidal and orbifold compactifications, see \cite{give89,give94} for
examples. Since the duality group can be embedded into the symplectic group
then (\ref{eq:comp-riem}) represents the transformation rule of
$\Omega\,(t, A_i)$ under T-duality.

There are also some more striking similarities between period matrices of
the three-form of Calabi-Yau manifolds and period matrices of Riemann
surfaces.  These are given by the appearance of the Theta functions that we
came across during the computations of the automorphic function. In fact
Riemann's theta functions, as defined before in (\ref{eq:riem-theta}),
\textit{are} defined within the context of Riemann surfaces.

In the context of a compact Riemann surface $C$, the $2g$ column-vectors of
the period matrix $\Pi =(\mathbb{1}, \Omega)$ generate a lattice $\Lambda$
in $\mathbb{C}^g$. The quotient space $\mathbb{A}\equiv
\mathbb{C}^g/\Lambda =\mathbb{C}^g/(\mathbb{Z}^g\Omega + \mathbb{Z}^g)$
defines a complex torus. Such torus is called the Jacobian variety of $C$,
denoted $J(C)$. Moreover, since the period matrix satisfies, 
\begin{equation}
  \Pi \left( 
    \begin{array}{cc}
      0 & \mathbb{1}\\ -\mathbb{1} &0
    \end{array} \right) \Pi^t \, = \, 0
\end{equation}

\noindent the Jacobian variety $\mathbb{A}$ is in fact a principally
polarized Abelian variety (rigorously, a principally polarized Abelian
variety is a pair consisting of a complex torus $\mathbb{A}$ together with
the Chern class $\chi$ of an ample line bundle on $\mathbb{A}$ such that
$1/g! \,\, \int_{\mathbb{A}} \chi =1$, see references
\cite{geometry,number,abelian}).

A \textit{Divisor} in $C$ is a linear combination of points, i.e.  $D
\equiv \sum_i n_i p_i$ for $p_i \in C, \,\, n_i \in \mathbb{Z}$.  Define a
meromorphic function $\mathrm{\phi}$ in $C$; then, in a local holomorphic
coordinate $z$, $h(z)$ defined by $\mathrm{\phi} = h(z)$ is a meromorphic
function. If $P \in C$ corresponds to the origin of the $z$-complex plane,
we can write $h(z) = z^\mu g(z)$. We define the \textit{order} of
$\mathrm{\phi}$ at $P$ as $\mu_P(\mathrm{\phi}) = \mu$.  The divisor
$(\mathrm{\phi})$ associated with the meromorphic function $\mathrm{\phi}$
is defined as $(\mathrm{\phi}) = \sum_{P \in C} \mu_P(\mathrm{\phi}) P$.
Due to the periodicity conditions of Riemann's theta function, defined by
(\ref{eq:theta-inv}) and (\ref{eq:theta-inv1}), over the lattice $\Lambda$,
the divisor $(\theta)$ induces a divisor on
$\mathbb{A}=\mathbb{C}^g/\Lambda$.  

Two Abelian varieties are isomorphic if their period matrices are related
by the symplectic transformation (\ref{eq:comp-riem}).  The moduli space of
the complex torus is therefore identified as the quotient space
$\mathscr{A}_g = \mathfrak{H}_g/Sp(2g, \mathbb{Z})$. Moreover, duality
transformations are induced by a particular subgroup of $Sp(2g,
\mathbb{Z})$, that given by the embedding of $SU(1,n)$ in the symplectic
group as seen in \cite{mine2}.

The action of the discrete group $Sp(2g, \mathbb{Z})$ on $\mathscr{A}_g$ is
holomorphic and well defined. Thus $\mathscr{A}_g$ is endowed with the
structure of a complex analytic space (a complex manifold away from 
some points fixed under the action of subgroups of $Sp(2g, \mathbb{Z})$).
One can define for this space a projective embedding $\mathscr{A}_g
\rightarrow \mathbb{P}^N \mathbb{C}$ for some $N \in \mathbb{Z}$. To define
this embedding one needs modular forms, in this case the so-called Siegel
modular forms, i.e. functions of $\Omega$ which transform as
\begin{equation}
  f \,[( U \Omega + V) \, ( W \Omega +Z)^{-1}] \, =
    \, \det( ( W \Omega +Z)^{-1})^k \, f (\Omega) 
\end{equation}
where $k \in \mathbb{Z}$ and 
$$\left(\begin{array}{cc} U & V \\ W & Z  \end{array}\right) \in \Gamma
\subset Sp(2g, \mathbb{Z}).$$
\vspace{1mm}

Riemann's theta function, as defined in (\ref{eq:riem-theta}), for example,
obeys the relation
\begin{equation}
  \theta (0, -\Omega^{-1})^2 = \det (-i\Omega)\,\,\, \theta(0,\Omega)^2.
\end{equation}
Hence, $\theta (0, -\Omega^{-1})^2$ is a modular form of weight 1.  In the
case $g=1$ that we addressed when we studied the space $SU(1,1)/U(1)$ the
automorphic function could be given in terms of a well knows modular
function (the Dedekind eta function) obtained from the Mellin
transformation of a theta function (via the Kronecker Limit formula).
However, showing that the automorphic function, in the general case, is a
Siegel modular function seems like a formidable task, at least from the
explicit expression (\ref{eq:finalmente}).

Nevertheless, the integral from which we originate our results, namely
(\ref{eq:limit}), is invariant under duality transformations. The integral
is regular up to a pole that we subtract, \textit{independent} of the
moduli and, therefore, the final result should keep these invariances.
In this sense one should be able to express the resulting function
(\ref{eq:finalmente}), in terms of Siegel modular forms. The correct
functional dependency is still work in progress.  Siegel forms are
presented in  \cite{nille97}.  It is interesting to note that
the threshold corrections to gauge couplings in $(0,2)$ compactifications
have been constructed in terms of the these modular forms \cite{mayr95}.
This type of compactification arise with the introduction of Wilson lines.

Another possible use of regularization procedure is the calculation of
integrals like those appearing in \cite{mayr95}, the integral proposed in
this reference is very similar to the integrals that we regularized here,
namely 
\begin{equation}
  \Delta = \int \frac{d \tau}{\tau_2} \sum_{k_1,k_2 \in \mathbb{Z}}
           \mathbf{Z}^{4d}(\tau,T,\mathcal{A},k_1,k_2)
           \mathscr{C}_a(\tau,k_1,k_2) 
\end{equation}
where $\mathscr{C}_a$ is a holomorphic moduli-independent function and 
\begin{eqnarray}
  \mathbf{Z}^{4d}& = & \! \! \! \! \sum_{n_1,n_2,m_1,m_2
    \in     \mathbb{Z}}  \! \! \! \!
    \exp\, \Big[ \frac{-\pi \tau_2}{\mbox{Im}(T -
    i\sqrt{3}/8|\mathcal{A}|^2)\, \mbox{Im}( U) } 
    \nonumber \\[5pt] 
 & &  \quad  \times |T U n_2 + T n_1 - U (m_1
     +\mathcal{A} k_1 - 1/2\, \mathcal{A} k_2) + m_2 - 1/2\, \mathcal{A}
     k_1|^2 \,\,      \nonumber \\ 
 & & \qquad \qquad \qquad\qquad \qquad \qquad + 
     \,2\pi i\tau (m_1n_1 +m_2  n_2) \Big].
\end{eqnarray}
Although not exactly the same problem, one can easily see how to use what
we have learned in this paper to compute this type of integral.

Going back a bit in our discussion one can conclude that there is a
correspondence between the moduli space of a Calabi-Yau manifold and the
moduli space of a Complex Abelian variety.  Moving on the moduli space of
the Calabi-Yau would correspond to moving on the moduli space of the
associated complex torus (associated through the period matrix $\Omega$).
Exactly how far this similarity between these two spaces can go is
difficult to establish. The structure of $\mathfrak{A}_g$ is still the
matter of much research by mathematicians, this is known as the ``Schottky
problem'' (see the references given above in this subsection).

The map $ (\tilde{z}_a, \, \tilde{\mathscr{F}}_a) \rightarrow
\Omega(\tilde{z}_a, \tilde{\mathscr{F}}_a)$ of (\ref{eq:vec-period}) and
(\ref{eq:comp-riem}), provides an isomorphism (at least locally)
\begin{equation}
  \frac{SU(1,n)}{U(1) \times SU(n)} \longrightarrow
    \frac{\mathbb{C}^{g}}{\mathbb{Z}^g \, \Omega + \mathbb{Z}^g} 
\end{equation}
where $2n = g = 2 b_{21}$ ($n$, the number of complex moduli, $g$ the genus
of the associated Riemann surface and $b_{21}$ the dimension of the
homology group respectively). Duality transformations in the compactified
string correspond to usual $SL(2,\mathbb{Z})$ transformations of the torus,
only generalized to the torus of given by the previous expression.  The
automorphic functions therefore are nothing but generalizations of the
usual elliptic functions usually associated with the torus. However, the
role of this underlying compact Riemann surface is not clear at the moment.

\section{Conclusions}
\markright{\thesection. Conclusions.}

The main result of this paper is the construction of automorphic functions
of manifolds with special K\"ahler geometry. This kind of geometry appears
in manifolds which are the moduli space of Calabi-Yau spaces. Our method
starts with the duality-invariant formula \ref{eq:master} and using
$\zeta$-functions we first obtain a well known result for the simple case
of $SU(1,1) / U(1)$, then we proceed to solve the case of $SU(1,2)
/ [SU(2) \times U(1)]$. 

It is also interesting to note that the construction of such solution is
one comes across Riemann theta functions which have a period matrix. Given
this plus the action of the symplectic group over this functions (which can
be seen as duality-transformations) one can interpret this matrix as the
period matrix of some Abelian Variety. Unfortunately, at this point, one
can only speculate about the nature of this underlying variety. This
situations is somehow similar to the Seiberg-Witten solution of $N=2$
supergravity and some research in this line could provide some interesting
insights into the structure of Calabi-Yau compactifications of string
theory.  Of course this connection has been noticed before by Candelas but
we are able to provide particular examples of such varieties. Generally
speaking our method can be generalized to other manifolds to which the
formula \ref{eq:master} applies. In many ways therefore one should be able
to reproduce the results of \cite{ober99}, \cite{kiri97} and others based
on the $SO(2,2n)/ [SO(2) \times SO(2n)]$ coset manifolds.

\paragraph{Acknowledgements.}  I have benefited from many useful
discussions with Dr. S. Thomas and Dr. Jos\'e Figueroa-O'Farrill for which
am deeply grateful; they not only gave me useful input but also read the
manuscript. I should acknowledge talks and encouragement from the people of
the Theory Group at Queen Mary \& Westfield College, London, as well.


\end{document}